\documentclass[prb,aps,superscriptaddress,showpacs,showkeys,reprint]{revtex4-1}
\usepackage{amsmath,amsfonts,amssymb}
\usepackage{bm}
\usepackage{natbib}
\usepackage{graphicx}
\usepackage{array}
\usepackage{verbatim}
\usepackage{graphicx,subfigure}
\usepackage{multirow}
\usepackage{url}
\usepackage{tikz}
\usepackage{hyperref}
\hypersetup{
    colorlinks,
    citecolor=red,
    filecolor=blue,
    linkcolor=blue,
    urlcolor=black
}

\usepackage[mathscr]{euscript}

\def\be{\begin{equation}}
\def\ee{\end{equation}}
\def\bea{\begin{eqnarray}}
\def\eea{\end{eqnarray}}
\def\ba{\begin{array}}
\def\ea{\end{array}}

\begin{document}
\preprint{APS/123-QED}
\title{Interference effect of  beam splitter current in Iron-Pnictide superconductors}
\author{Abhisek Bag}
 \altaffiliation[]{abhisek.b@iopb.res.in}
\author{Saptarshi Mandal}%
 \email{saptarshi@iopb.res.in}
\affiliation{Institute of physics, Bhubaneswar 751005, Odisha, India}
\affiliation{Homi Bhabha National Institute, Mumbai 400 094, Maharashtra, India}
\date{\today}
\begin{abstract}
We consider a Cooper pair beam splitter for Iron-Pnictide  superconductor and calculate the entangled
electron-hole current. We investigate the interplay of various physical parameters such as doping at electron and hole pockets as well as non-zero nesting between the electron and hole pocket. We find that in the absence of magnetic order, the current due to hole pocket and electron pocket  add up ordinarily. However in the presence of magnetic ordering the two currents take part in characteristic  interference effect to modify the resultant current significantly. This interference effect manifests itself in  non-monotonous and oscillatory nature of beam splitter current. We investigate in details this non-monotonicity with the chemical potential as well as  nesting vectror $|\bf q|$. We also investigate the evolution of density of states  with system parameters and correlate it with the beam-splitter current. Further  we  enumerate the relevant parameter space where the efficiency of such beam splitter  set up is enhanced. Our finding can be useful in experimental determination or verification of co-existence phase in Iron-Pnictide superconductors and has potential applications in realizing quantum gates or switches.  
\end{abstract}
\maketitle
\section{\label{sec:level1}Introduction}
 One of the fundamental  aspect of quantum mechanics is the superposition principle which dictates that a quantum mechanical system can be described by a linear superposition of many  orthogonal wave functions \cite{sakurai}. This superposition principle yields an entanglement or correlations between different parts of the system. In recent times this fundamental aspect has been  investigated in detail to construct entangled states having  specific properties with application in teleportation, cryptology etc \cite{neilson}. To realize such entangled states many protocols  have been proposed  involving  various  mesoscopic system such as quantum dots, superconducting qubits etc \cite{divincenzo,djordjevic,gaard,montanaro}. However after preparing such  entangled state one needs to confirm that the state is robust against  decoherence \cite{decoher1,decoher2}. It is  well known fact that  various condensed matter system  act as a source of such entangled many body states naturally. The existence of such entangled states is ubiquitous in various condensed matter systems such as the Cooper pairs in superconductivity \cite{tinkham}, ground state wave-function  for quantum spin liquid \cite{spinliq1,spinliq2}, resonating valence bond states \cite{baskaran} for certain magnetic systems etc.

\indent
Starting from the celebrated EPR paradox \cite{Einstein} and culminating in Bell's inequality\cite{bell,peres},  entanglement is now an experimentally established reality\cite{wen,nicolas,osterloh}. In view of identifying  such entangled states in the context of superconductivity, a Cooper pair beam splitter arrangement has been proposed for the first time in  \onlinecite{daniel} and later on in other systems \cite{hofstetter2,herrmann,burset-2011,sato-2014,mantovani-2019,valagian-2019}. In these study it is proposed that  the participating electrons and holes  can be collected at two different leads after passing through two quantum dots kept at  suitable distance. The collected electron and hole maintain their state of initial correlation  or entanglement. 
Recently such  studies has been extended to graphene \cite{firoz} where a proximity induced superconductivity has been considered. There the authors have shown that the  beam  splitter current is increased in magnitude in comparison to an ideal BCS superconductor in two dimensions.  However the superconductivity in single layer graphene is not realized  till now and the superconductivity in conventional BCS superconductor is realized at very low temperature limiting its  practical application. Thus it motivates us to look for other system where such entangled beam splitter can be realized with naturally occurring unconventional superconductivity \cite{norman,bang-2017,singh-2008,wu-1987}  at high temperatures enabling us for practical applications.\\
\indent
In this study we have investigated the consequences of beam splitter arrangement for Iron-pnictide superconductors. Firstly we are motivated by the fact that this system is an example where  various competing order parameters such as two superconducting pairing interactions and  magnetic ordering  exist simultaneously \cite{hhwen,stewart,hosono,qsi,steglich-2007}. Till date all the beam-splitter superconducting current studies are performed on systems containing only single pairing interaction. Thus consequences of many order parameters render it an interesting platform to study beam-splitter current. Secondly these systems are realized at quite high temperature for example at  $\sim$50K for 1111- and 122- type systems \cite{hosono} thus providing relatively easier practical realisation. In fact we have found that the Cooper pair beam splitter current depends on the  relative magnitude of chemical potential, two superconducting gaps and the magnetic order parameter in a nontrivial way. Though the exact expression for beam splitter current can not be obtained in the simultaneous presence of magnetic and superconducting order parameter like previous studies, our study offers numerous valuable  consequences of the complexities of the system on beam-splitter current. We have also seen that unlike before, the beam-splitter current can be monotonically decreasing or an oscillating one depending on the relative  shapes of the electron and hole pocket (associated with two superconducting gaps) as well as  values of chemical potential at electron  and hole pocket respectively.  We attribute this to the interference effect from currents due to hole and electron pocket.  We also investigate the effect of the density of states of the system on the beam splitter current. \\
\indent
Our plan of presentation is the following.   In section \ref{citeref} we review the  basics of Iron-pnictide superconductors \cite{singh-2012,zhangh-2008,chubukov-2009}, its model hamiltonian involving electron and hole pocket and brief description of superconducting and magnetic interaction between the hole and electron pocket in the system . After this, in section \ref{iron-beam-split}, we present our  model scheme of realising  beam-splitter current in detail by calculating the  entangled beam-splitter current in its full form for the Iron-pnictide system. We also enumerate all the assumptions and approximations needed for such calculation.  In section \ref{results}, we  first present  the phase diagram of model hamiltonian discussed and the relevant range of parameters  considered  in the present study to explore the nature of beam splitter current. This includes all the relevant parts of the phase diagram. We present the beam-splitter current in the phase space as a whole
at first. Then we discuss  how  beam splitter current varries with  the  distance between the two leads for  various representative values of the parameters of the system such as chemical potential at electron and hole pocket, perfect and non-zero nesting etc. We also present the study of density of states and correlate it with the beam-splitter current. In Sec. \ref{i2} we briefly discuss the current when both the electrons are collected at the same dot and comment on the efficiency of the entangled beam-splitter current. Finally we conclude in section \ref{discussion} by summarising our results  with a discussion.
\section{Model}
\label{citeref}
In FeAs, the relevant tight binding  model which describes the superconductivity contains an itinerant electron system of two electronic orbitals due to two Fe atoms in an unit cell~\cite{araujo,bang,hu-ting-zhu,jiang-li-wang}. The hybridization between the orbitals are such that it leads to a hole pocket centred at $(0,0)$ and an electron pocket centred at $(\pi, \pi)$ in the folded zone scheme \cite{chubukov1,chubukov2}.  Following earlier notations\cite{chubukov1}  we denote the fermions near $(0,0)$ by usual `$c$' operator and for fermions near  $(\pi,\pi)$ by `$f$'.  The model Hamiltonian can be written as follows,
\begin{equation}
        H=H_{0}+H_{\Delta}+H_{m} .
\end{equation}
In the above $H_0$  denotes the  non-interacting part  of electron and hole pocket. $H_{\Delta}$ denotes superconducting pairing interaction and $H_{m}$ signifies  the magnetic pairing interaction. The details of these three terms are given  below \cite{chubukov2},
\begin{eqnarray}
&&H_{0}=\sum_{k}\epsilon_{c}(k)c_{k\alpha}^{+}c_{k\alpha}+\sum_{k'}\epsilon_{f}(k')f_{k'\alpha}^{+}f_{k'\alpha} 
\label{ekcf}
\end{eqnarray}

\begin{figure}[h!]
\label{setup1}
\includegraphics[height=4cm,width=10cm]{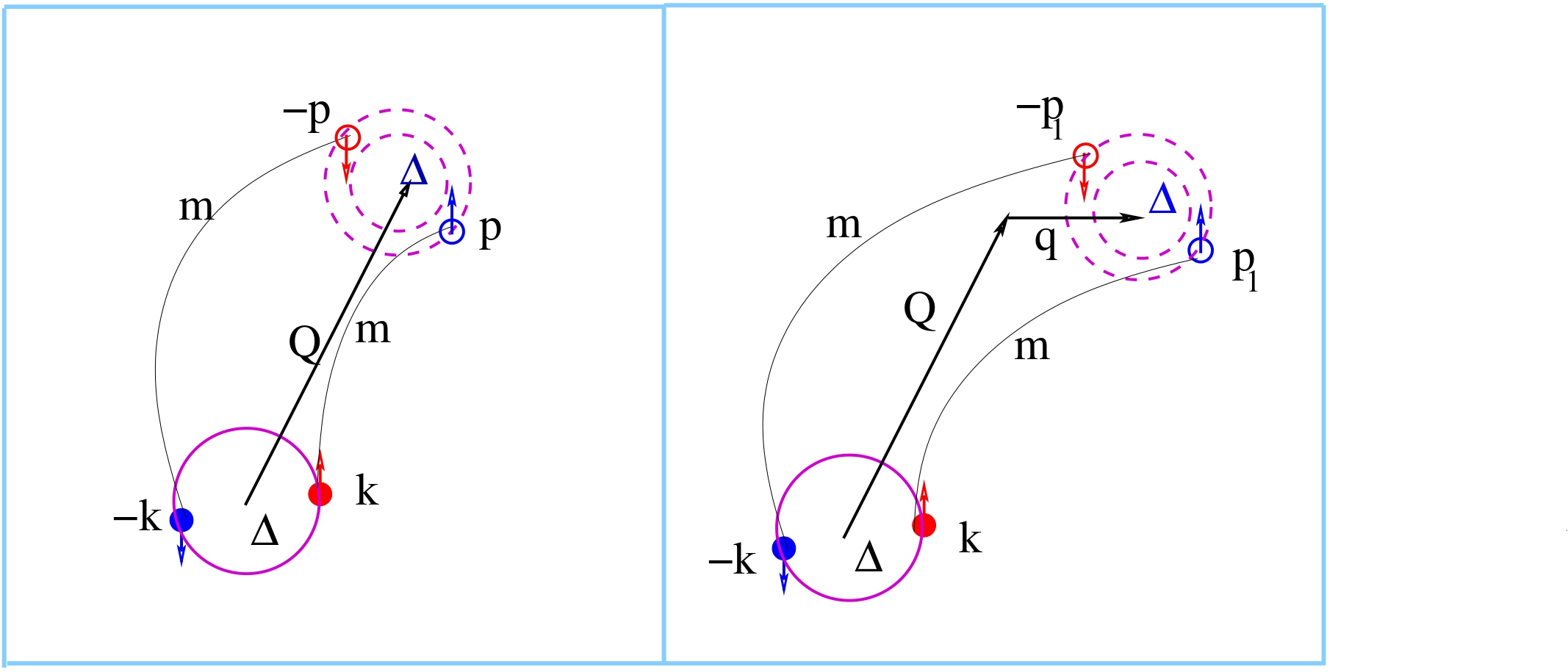}
\caption{The solid circle denotes hole pocket and the dashed circles denote electron pocket. The black lines connecting  fermions at electron and hole pocket are responsible for magnetic ordering in the system. The left panel shows the case of perfect nesting and the right panel shows a non-zero nesting (finite  {\bf q}). $\Delta$ denotes the superconducting gap at the electron and hole pocket. The effect of chemical potential increases or decreases the relative size of electron and hole pocket which have been shown by two different dashed circles. }
\end{figure}
In the above $\epsilon_c= \mu_c - |k|^2/2m$  and $\epsilon_f = |k^{\prime}|^2/2m - \mu_f$. Note that $k$  appearing in `$c$' is measured from the center of Brillouine zone and  for `$f$' fermion $k'$ is measured from  ($\pi,\pi$). Henceforth all the momentum indices associated 
with the '$f$' fermions are understood as measured from ($\pi, \pi$) unless referred otherwise. Note that we have taken an isotropic mass at electron pocket and  hole pocket. This assumption may be an approximated one without changing the qualitative aspect of the results found here.
The superconducting pairing terms  and the magnetic interactions are  given by,
\begin{eqnarray}
\label{delta}
H_{\Delta}&=&\frac{1}{2}\sum_{k,p}V_{[\Lambda]}^{cf}(k,p)(c_{k\alpha}^{+}c_{-k\beta}^{+}f_{-p\beta'}f_{p\alpha'} + h.c)~~~~ \\
\label{mag}
H_{m}&=&-\frac{1}{4}\sum_{\delta p=\delta k}V_{[\Lambda]}^{sw}(p'p:kk')(f_{p'\alpha}^{+}c_{p\beta}c^{+}_{k\beta'}f_{k'\alpha'} + h.c )~~~~
\label{mag}
\end{eqnarray}
Where $\delta p= p'-p$ and $\delta k= k'-k$ and $[\Lambda]=\alpha\beta\beta^{\prime}\alpha^{\prime}$ denotes combined spin indices. In the above, Eq. \ref{delta} describes the effective pairing interactions in the Iron-pnictide system which arises due to the inter-band  hopping between electron and hole band \cite{chubukov1}. The magnetic interactions as denoted in Eq. \ref{mag} come from the inter-band density-density interactions.  The pairing interaction $ V_{\alpha\beta\beta'\alpha'}^{cf}(k,p)=V_{k,p}^{sc}((i\sigma^{y})_{\alpha\beta}(i\sigma^{y})^{+}_{\beta'\alpha'}$ and the magnetic interaction $ V_{\alpha\beta\beta'\alpha'}^{sw}(p'p:kk')=V_{p'p:kk'}^{sw}\boldsymbol{\sigma}_{\alpha\beta}.\boldsymbol{\sigma}_{\beta'\alpha'}^{+}$. 
Following earlier convention \cite{chubukov1,chubukov2}, we define the  superconducting pairing order parameter and magnetic order parameter as $\Delta_{c}(k)_{\alpha\beta}=(i\sigma^{y})_{\alpha\beta}\sum_{p}V_{kp}^{sc}(i\sigma^{y})_{\beta'\alpha'}^{+}\langle f_{-p\beta'}f_{p\alpha'} \rangle, (m_{q})_{\alpha\beta}=-\frac{V^{sw}}{2}\sum_{p}\boldsymbol{\sigma_{\alpha\beta}}.\boldsymbol{\sigma_{\beta'\alpha'}^{+}}\langle c_{p\beta'}^{+}f_{p+q\alpha'}\rangle $. $\Delta_f(k)_{\alpha, \beta}$ is defined identically as $\Delta_c(k){\alpha, \beta}$. Note that the momentum index $q$ appearing in the definition of magnetic order parameter $m_q$ refers a small deviation of the centre of hole pocket from ($\pi,\pi$). This is to account for the fact that when $q=0$, magnetic order is found to be zero which will be discussed again Sec. \ref{results}.
We use the above definitions of mean field order parameters to perform a  meanfield decomposition of the four fermion interactions as given in Eq. \ref{delta} and Eq. \ref{mag}. The resulting Hamiltonian obtained is written in the following  matrix form,

\begin{eqnarray}
\label{finmom}
H&=&\sum_{k\alpha\beta}\overline{\Psi}_{k\alpha}\hat{\mathcal{H}}(k)_{\alpha\beta}\Psi_{k\beta}
-2\frac{\Delta_{c}\Delta_{f}}{V_{SC}} +2\frac{m_{\alpha\beta}^{*}m_{\alpha\beta}}{V_{SD}},
\end{eqnarray}
where the four component Nambu spinor is,  $\overline{\Psi}_{k\alpha}=(c_{k\alpha}^{+},c_{-k\alpha},f_{p\alpha}^{+},f_{-p\alpha}) $ with $p=k+ q$ and the Hamiltonian  matrix $\hat{\mathcal{H}}(k)_{\alpha \beta}$  which we write in short $\hat{\mathcal{H}}_{\alpha \beta}$ has the form,\\
\begin{eqnarray}
\label{4matrix}
\hat{\mathcal{H}}_{\alpha\beta} = \begin{pmatrix} 
\epsilon_{c}(k)_{\alpha\beta} & \Delta_{c}(k)_{\alpha\beta} & m^{*}_{q\alpha\beta} & 0 \\
-\Delta_{c}(k)_{\alpha\beta} & -\epsilon_{c}(\tilde{k})_{\alpha\beta} & 0 & -m^{*}_{q\alpha\beta}  \\
m_{q\alpha\beta} & 0 & \epsilon_{f}(p)_{\alpha\beta} & \Delta_{f}(p)_{\alpha\beta} \\
0 & -m_{q\alpha\beta} & -\Delta_{f}(p)_{\alpha\beta} & -\epsilon_{f}(\tilde{p})_{\alpha\beta} 
\end{pmatrix}~~~
\end{eqnarray}

 Note that initially the superconducting pairing interaction (as given in Eq. \ref{delta}) is defined such a way that a hole of momenta $k$  can interact with another electron of momenta $p$ belonging to electron pocket centred around $(\pi,-\pi)$. However in the final expression in Eq. \ref{finmom} another simplification has been performed where a hole with momenta $k$ take part in pairing interaction with another electron at hole pocket with momenta $k$ only. The index $q$ signifies that the electron pocket is no longer around $(\pi, -\pi)$ but rather around $(\pi,-\pi) + q$ to take into account finite magnetic order parameter. In the above we have accounted for  two different kind of magnetic order parameter with the  following definitions. If $\alpha=\beta,  m_{\alpha,\alpha}=m_1$ and  for $\alpha \neq \beta,~~ m_2= m_{\alpha, \beta}$.
Following earlier study \cite{chubukov1,chubukov2} we choose $m_1=m, m_2=0$ in our study without loss of generality. One can easily obtain the spectrum after diagonalizing the $8 \times 8$ matrix in Eq. \ref{4matrix}, the detail expressions for the eigenvalues are given in Appendix B. We obtain the ground state energy by filling the negative energy eigenstates which correspond to  the vacuum  of new Bogoliubov quasi-particles. Along with the constant terms the ground state energy can be written as $E_{min}=- \mathcal{E}_{kq+} - \mathcal{E}_{kq-} -\frac{\Delta_c \Delta_f}{V_{SC}}+2\frac{m^2}{V_{SD}}$.
Where $\mathcal{E}_{kq\pm}= \sqrt{\mathcal{P}_k \pm \sqrt{\mathcal{Q}_k}}$ where $\mathcal{P}_k$ and $\mathcal{Q}_k$ are functional of $\epsilon_c(k), \epsilon_f(p), \Delta_c, \Delta_f, m$ and their detail expressions are given in Appendix B.  We first derive the self consistent equations for  the meanfield order parameter by minimizing Eq. \ref{finmom} with respect to appropriate meanfield order parameters. Next we evaluate those meanfield order parameters self-consistently  with the help of eigenvectors obtained after diagonalizing Eq. \ref{4matrix}.  Once the meanfield order parameters values are obtained we use them to calculate  the beam-splitter current. Note that we evaluate the meanfield order parameters at zero temperature and thus beam-splitter current refers to zero temperature study only.  Now we move on briefly to describe the beam-splitter current calculation. 
\section{Model for Beam Splitter Geometry}
\label{iron-beam-split}
\begin{figure}[h!]
\includegraphics[height=4cm,width=8cm]{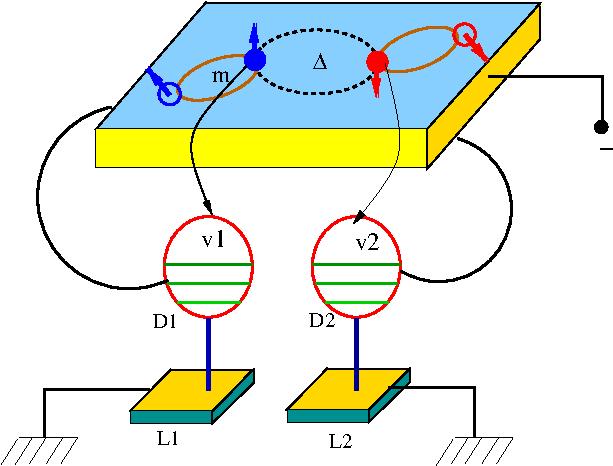}
\caption{\label{setup}Cartoon picture of Cooper Pair Beam Splitter set-up for Iron-pnictide superconductor. The blue rectangular slab with yellow border represents the superconductor and a Bogoliubov quasiparticle has been drawn schematically on it. The two filled circles represent fermions participating in superconducting order parameter $\Delta$ and the magnetic order is represented by the coupling between one filled circle and one empty circle. The superconducting slab is connected to two quantum dots (D1 and D2) by black lines. Each quantum dot is finally connected to fermi ( metallic) leads L$_1$ or L$_2$. The direction of the current is indicated by the green arrows. }
\end{figure}

Having described the basic model of superconductivity and coexistence of magnetic order with it in Iron-Pnictides we now elaborate on the beam splitter arrangement designed for such system. Our idea of entangled Cooper pair splitter is similar to the seminal work of \cite{daniel} where for the first time an outline of Cooper pair entangled beam splitter arrangement was introduced. In Fig. \ref{setup}, we present a schematic diagram of entangled beam-splitter arrangement. It is expected that electrons in a superconducting Copper pair in an iron-pnictide material will be tunneling through two spatially separated quantum dots where each quantum dot  allows one electron (among the two participating electrons in a superconducting pair).  However we note that in previous studies \cite{daniel,firoz} a superconducting eigenstate is of the form $\gamma_k = u_k c_{\uparrow, k} + v_k c^{\dagger}_{\downarrow, -k}$ and hence there are possibilities of only two kind of currents. The first one is where the electron with spin up and momentum $k$ enters into one of the quantum dot and the other enters into the remaining quantum dot. This current is defined as $\mathcal{I}_1$ and it shows characteristic oscillations with the separation between the quantum dots. The second kind of current is obtained where both the electrons enter through the same dot with some time gap in between them. This current is defined as $\mathcal{I}_2$ and it is found to be a constant for a given set of  parameters. The ratio $\mathcal{I}_1/\mathcal{I}_2$ signifies the efficiency of the beam splitter current so as to differentiate the $\mathcal{I}_1$ from $\mathcal{I}_2$. In the present scannerio an eigenstate in superconducting phase is of the form $\gamma^{\prime}_k=u_k c_{\uparrow, k} + v_k c^{\dagger}_{\downarrow, -k} + u^{\prime}_k f_{\uparrow, k} + v^{\prime}_p f^{\dagger}_{\downarrow, -p} $. Hence in our case the current $\mathcal{I}_1$ is obtained  by considering the two participating entangled Cooper pair, one due to $c$-fermion and another due to $f$-fermion. Note that we do not consider the current due to tunneling of a $c$ and $f$ fermion which participate in magnetic ordering. The current due to such event vanishes due to singlet nature of the quantum dot and metallic lead. \\
\indent

  The current $\mathcal{I}_1$ is expected to happen once the quantum dots are kept in the Coulomb blockade regime such that it is  energetically  unfavorable to accommodate two electrons at the same quantum dot. Finally once two spatially separated  electrons enter into two separate quantum dots, they are collected by two fermi liquid leads (represented by L$_1$ and L$_2$ in Fig. \ref{setup}).  However to realize such entangled electrons to be separated the chemical potentials of the superconducting materials ($\mu_s$), quantum dots($\epsilon_1$ and $\epsilon_2$) and also the leads($\mu_{1}$ and $\mu_{2}$) are to be kept in certain conditions. While the chemical potentials at the two fermi liquid leads can be
kept equal generally, the chemical potentials of the two quantum dots should be such that   $\epsilon_1 + \epsilon_2=2 \mu_s$ corresponding to  two particle Breit-Wigner resonance \cite{Sumetski}. The transport of two entangled electron or hole pairs from the quantum dots to the leads can be achieved by applying a bias voltage $\Delta \mu= \mu_s - \mu_l$.   We here outline the calculation of $\mathcal{I}_1$ and the details about $\mathcal{I}_2$ is discussed in Sec. \ref{i2} and in  Appendix-A.\\
\indent
It is  natural to expect that once the electrons tunnel from the superconductor to quantum dots, they interact with the already existing electrons at the quantum dots and this might lead to  decoherence between the two separated electrons. One way to avoid this is to work in the co-tunneling regime  where the number of electrons in the quantum dots are fixed and also
the resonant level $\epsilon_1,~\epsilon_2$ are not occupied. The probability of interaction can also be reduced if
the entered electrons spend less time in the quantum dots and this can be achieved by having $|T_{SD}| < |T_{DL}|$. 
Also the temperature of the superconductor and the quantum dots should be such that $\Delta{\mu} > K_B T$ as this will ensure that the stationary occupation due to the coupling to leads is exponentially small.
\indent
The complete Hamiltonian for the proposed  beam splitter device thus contains the intrinsic Hamiltonian of the Iron-pnictide materials, two quantum dots and two fermi (metallic) liquid quantum leads. In addition to the intrinsic Hamiltonians mentioned above, it also involves  the  tunneling Hamiltonians between the superconductor and the quantum dots and between the quantum dots and the leads.  The complete Hamiltonian represented by $H_{Tot}$ for the beam splitter set up thus becomes,
\begin{eqnarray}
 H_{tot}&&=H_S+  \sum_l H_{D,l} + \sum_l H_{L,l}+ H_{SD} + H_{DL} .~~~~~~
 \label{comham}
\end{eqnarray}

In the above the first three terms indicate the intrinsic Hamiltonians of superconducting material, quantum dots and lead respectively.  Henceforth the subscript $S, ~D,~ L$ will be used for the superconductor, quantum dots and the leads respectively.
`$l$' denotes the number of leads and quantum dots  which can take values 1 and 2.
The term $H_{SD}$ and $H_{DL}$ denote the tunneling Hamiltonians from superconductor to quantum dots and quantum dots to leads respectively.  Below we give in detail the complete expressions of each terms in Eq. \ref{comham}. 
\begin{eqnarray}
\label{detail-ham}
H_{S}=&&\sum_{k\sigma}[\mathcal{E}_{kq+} \gamma^{c+}_{kq\sigma}\gamma^{c}_{kq\sigma}+\mathcal{E}_{kq-}\gamma^{f+}_{kq\sigma}\gamma^{f}_{kq\sigma}] + \mathcal{E}_0\\
H_{Dl}=&&\sum_{\sigma,l}\epsilon_{l} d_{l\sigma}^{+}d_{l\sigma}+Un_{l\sigma}n_{l-\sigma}, \\
H_{Ll}=&&\sum_{k\sigma}\epsilon_{k}a_{lk\sigma}^{+}a_{lk\sigma}\\
H_{SD}=&&\sum_{l\sigma,i}{g_i}T_{SD}^{i}d_{l\sigma}^{+}\Psi_{\sigma}^{i}(r_{l}) + h.c \\
H_{DL}=&&\sum_{lk\sigma}T_{DL}a_{lk\sigma}^{+}d_{l\sigma} + h.c
\end{eqnarray}

In the above $\mathcal{E}_{kq\pm}$ denotes the energy of  Bogoliubov quasi-particle excitations discussed before.  $\gamma^{c}_{kq\sigma}$ and $\gamma^f_{kq\sigma}$ are creation operators for Bogoliubov quasi-particle. $\mathcal{E}_0$ denotes the constant energy not important to our consideration and does not affect the beam-splitter arrangements.  $\epsilon_l$ denotes the energy levels of quantum dots and $U$ refers the interaction energy if a given energy level of the quantum dot is occupied by two electrons of opposite spins. $\epsilon_k$ describes the  energy of the Bloch states of fermi liquid leads. $T^i_{SD}$ is the tunneling amplitude between superconductor and $i$'th quantum dots and $\Psi^i$ represents a state of an electron/hole arriving at $i$'th quantum dot. $g_1$ and $g_2$ 
can take value 0 or 1 depending on which kind of quasi-particle is taking part in the Andreev process. We note that $g_1 \neq g_2$ signifies  two types of quasi particles can not be present simultaneously in a pair of dots. Lastly $T_{DL}$ denotes the tunneling amplitude between quantum dots and fermi liquid leads. 
\subsection{\label{sec:level2}Current: via different dots}
Having provided a brief introduction on Iron-pnictide system in Sec. \ref{citeref} and the beam-splitter set up in detail in Sec. \ref{iron-beam-split}, we now  outline the important steps in  evaluating  beam-splitter current which are finally measured at the fermi liquid leads. The current  due  to electrons coming  from superconductor and finally reaching  to lead via quantum dots are represented by \cite{daniel} $I=2e\sum_{f,i}W_{f,i}\rho_{i} $
where $W_{f,i}=2\pi|\langle f|T_{\epsilon_{i}}|i \rangle|^2\delta(\epsilon_{f}-\epsilon_{i}) $ is the transition rate  from an initial state of electron $| i \rangle$  to a final state $\langle f|$ which is discussed below in detail  and $\rho_i$ denotes the electron density. Here $T(\epsilon_i)$ is the on-shell transmission matrix and is given by~\cite{Merzbacher} $T(\epsilon_{i})=H_{T}\frac{1}{(\epsilon_{i}+i\eta-H)}(\epsilon_{i}-H_{0})$.
In the above $H_0$ denotes the Hamiltonian for superconductor ($H_0= H_{S}$) and $H_T$ denotes the successive tunneling from superconductor to quantum dots and subsequently from quantum dots to leads. $H=H_C$, is the total Hamiltonian. As $H_T << H_0$, we can expand the denominator in appropriate power series and we obtain in $\eta \rightarrow 0$ limit,

\begin{eqnarray}
\label{totalT}
T(\epsilon_{i}) &&=H_{T}\frac{1}{(\epsilon_{i}+i\eta-H_0-H_T)}(\epsilon_{i}-H_0)\nonumber \\
  &&=H_{T}\frac{1}{\tilde{H}_0\Big( 1-\frac{H_{T}}{\tilde{H}_0} \Big)}(\epsilon_{i}-H_0),~~ \tilde{H}_0=\epsilon_{i}+i\eta-H_0  \nonumber \\
  &&=H_{T}\frac{1}{\tilde{H}_0}\Bigg( 1+\sum_{n}\Big(\frac{H_{T}}{\tilde{H}_0}\Big)^n \Bigg) (\epsilon_{i}-H_0) \nonumber \\
&& =H_{T}+H_{T}\sum^{\infty}_{n=1}\Big(\frac{H_{T}}{\epsilon_{i}+i\eta-H_{0}}\Big)^n
\end{eqnarray}
In the last step we have employed the fact that $\eta\to{0}\nonumber$ and summation over `$n$' runs from zero to infinity as indicated.
We note that $\langle f|T(\epsilon_{i})|i \rangle$ denotes two step processes such that $\langle f|T|i \rangle=\langle f|T'|f'\rangle \langle f'|T''|i \rangle$  where $T^{\prime \prime}$ denotes the tunneling from superconductor to quantum dots and $T^{\prime}$ denotes the tunneling from quantum dots to leads. The expressions for $T^{\prime}$ and $T^{\prime \prime}$ are obtained as\cite{daniel}, 
\begin{eqnarray}
\label{firstT}
&&T''=\frac{1}{i\eta-H_{0}}H_{SD}\frac{1}{i\eta-H_{0}}H_{SD} \\
\label{secondT}
&&T'=H_{DL}\sum^{\infty}_{n=0}\Bigg(\frac{H_{DL}}{i\eta-H_{0}}\Bigg)^{2n+1} .
\end{eqnarray}
In arriving at the above form of the tunneling amplitude we have considered the fact that $|T_{SD}| < |T_{Dl}|$ and while the electron can come and go from the dot to lead many times, there is no possibility of an electron to go back from quantum dot to superconductor~\cite{daniel}.
We note that  $|i \rangle=|0\rangle_{S}\otimes|0\rangle_{D} \otimes|\mu_{l}\rangle_{l} $ where the subscript $S, D, l$ are used to denote the initial states of superconductor, quantum dots and leads respectively. The  states $|f \rangle=|LL \rangle, | f^{\prime} \rangle= |DD\rangle$ denotes the final states of leads and quantum dots respectively which are due to arrival of electrons from superconductor and hence represents a state with higher number of electrons and can be  written  as,
\begin{eqnarray}
\label{finstate}
&&|f\rangle=\frac{1}{\sqrt{2}}(a_{1p\uparrow}^{+}a_{2q\downarrow}^{+}-a_{1p\downarrow}^{+}a_{2q\uparrow}^{+})|i\rangle,\\
\label{finstate1}
&&|f'\rangle=\frac{1}{\sqrt{2}}(d_{1\uparrow}^{+}d_{2\downarrow}^{+}-d_{1\downarrow}^{+}d_{2\uparrow}^{+})|i\rangle.
\end{eqnarray}

Taking into account the orthogonality of states of different occupation numbers and momentum conservation, we obtain $\langle f|T'|f' \rangle=\langle i|a_{2q\downarrow}a_{1p\uparrow}T'd_{1\uparrow}^{+}d_{2\downarrow}^{+}|i\rangle$
We now turn to simplify $ \langle f'|T''|i\rangle=\frac{1}{\sqrt{2}}\langle i|(d_{2\downarrow}d_{1\uparrow}-d_{2\uparrow}d_{1\downarrow})T''|i \rangle$, which denotes Andreev process. It may happen that in the process of transport one electron with a particular spin(say up) arrives  at the quantum dot from the superconductor but an electron with opposite spin(say down) may travel forward from the same dot to the lead but we forbid that process. We want processes such as  $ |SS\rangle\to|DS\rangle\to|DD\rangle $ and $|SS\rangle\to|SD\rangle\to|DD\rangle $ simultaneously. Such that entangled pair of electrons from the superconductor(SS) get transported to two dots (DD) and move forward to the respective leads, because we avoid spin flipping. Where, $|SD \rangle=\gamma_{k\sigma}^{+}d_{l-\sigma}^{+}|i \rangle$. We ensure  that  $H_{S_1D_1}$  selects one electron of the entangled pair to dot 1 and $H_{S2D2}$ selects the other electron of the entangled pair to dot 2. A little algebra gives us the following expression,
 \begin{eqnarray}
 \mathcal{I}_1 && = \mathcal{C}_0 \mathcal{I}_D,~~~\mathcal{I}_D=  \Big(\sum_{k}{\mathcal{A}_k\cos(k_f\delta{r})}\Big)^2
  \label{fincurrent}
 \end{eqnarray}
In the above $\delta r$ represents the seperation between two quantum dots. For detailed derivation of the above equation we refer Appendix-A.  In the above $\mathcal{C}_0= \frac{e\gamma_{s}^2\gamma}{4\pi^2\nu_{s}^2[(\epsilon_1+\epsilon_2)^2+(\gamma^2/4)]} $ and denotes a constant depended on system specification. Here $\gamma=\sum\gamma_{l}; l=1,2$ and $\gamma_{l} = 2\pi\nu_{l}{T_{DL}}^2$(we remind that $l=1,2$ denote leads). Also $\gamma_s = 2\pi\nu_{s}T_{SD}^2$, where $\nu_{s},\nu_{l}$ are density of states per spin at lead and superconductor respectively. To understand the consequences of the magnetic order parameter and superconducting pairing, we expand the expressions of $\mathcal{A}_k$ below. 
 
\begin{equation}
\label{formA}
\mathcal{A}_k = \sum^{\alpha=c,f}_{i=1,3} \frac{u^{\alpha}_{i,k}u^{\alpha}_{i+1,k}- u^{\alpha}_{i,-k}u^{\alpha}_{i+1,-k}}{E_{ikq}}
\end{equation}
 In the above we have used $E_{1kq}= \mathcal{E}_{kq+}$ and $E_{3kq}= \mathcal{E}_{kq-} $.
  We note that in the absence of magnetic ordering the first and third term vanishes and we are left with two copies of superconducting pockets. In the presence of magnetic ordering all the four terms are present and consequently the amplitude of the  beam-splitter current is modified significantly.
  As an electron or hole has to participate simultaneously to the formation of superconducting pairing as well as in magnetic ordering the amplitude of beam-splitter current is expected to be modified significantly. 
In the following, we investigate $\mathcal{I}_D$ instead of the full current $\mathcal{I}_1$ as in Eq. \ref{fincurrent} following earlier convention~\cite{firoz}, which anyway is different only by a constant factor. We evaluated $\mathcal{I}_D$ numerically and plotted against $k_f \delta{r}$  for different values of system parameters such as doping at electron and hole pocket, nesting vectors etc in the next section. In the following text we simply refer $\mathcal{I}_D$ as the beam-splitter current. We also note that all our calculations are done at zero temperature.
 
\section{Numerical Results }
\label{results}
\begin{figure}[h!]
\includegraphics[height=5.2cm,width=8.1cm]{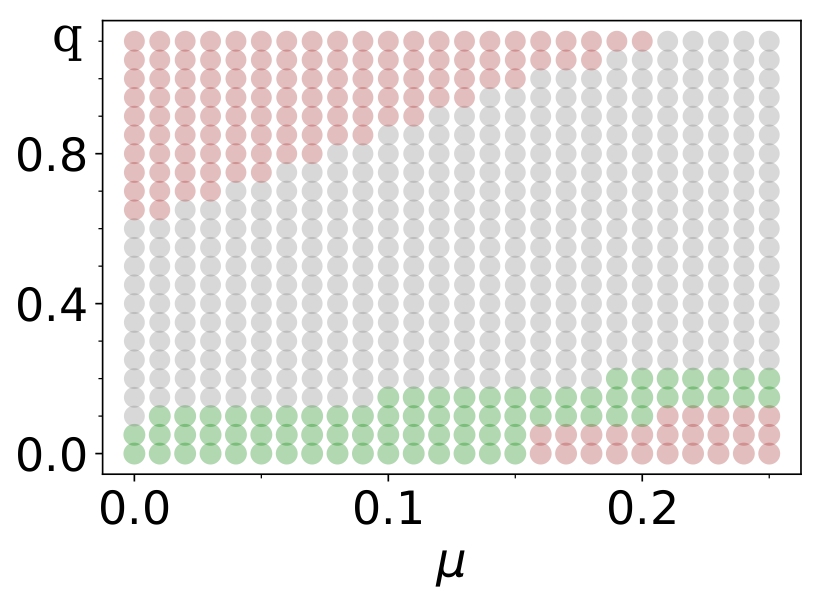}
\caption{\label{qmuphase}The phase diagram in {\bf q}-$\mu$ plane. The region filled with green filled circle denotes magnetic phase. The pink colored region denotes superconducting phase and grey colored region is the co-existence phase where superconducting and magnetic order co-exist. }
\end{figure}
\begin{figure}[!tbp]
    \centering
    \subfigure[]{\includegraphics[width=0.23\textwidth]{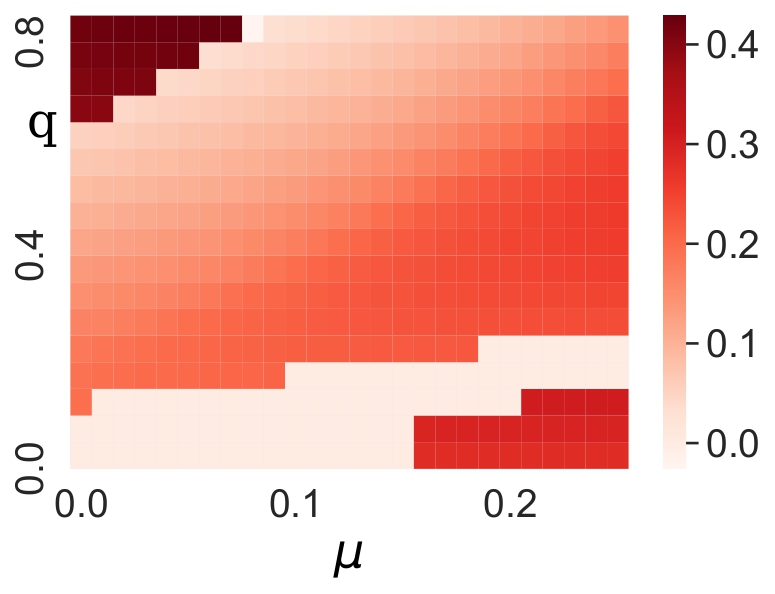}}
    \subfigure[]{\includegraphics[width=0.23\textwidth]{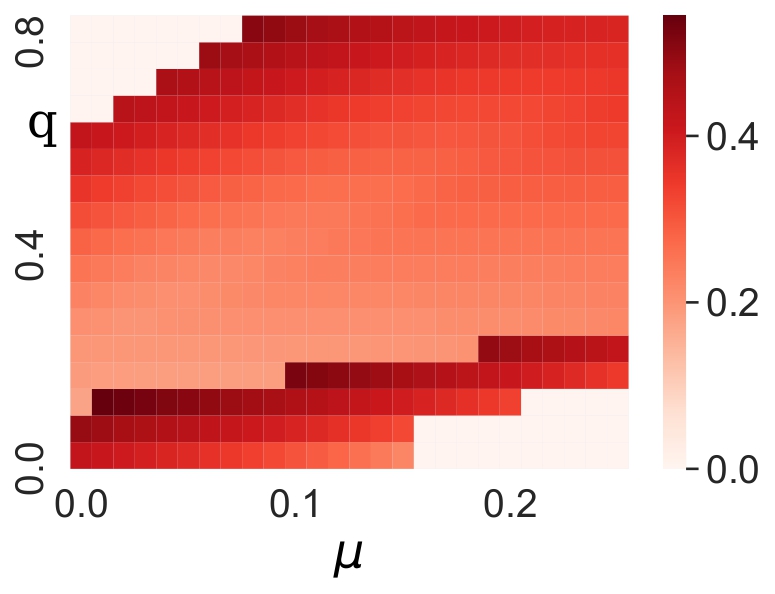}}\\
    \subfigure[]{\includegraphics[width=0.23\textwidth]{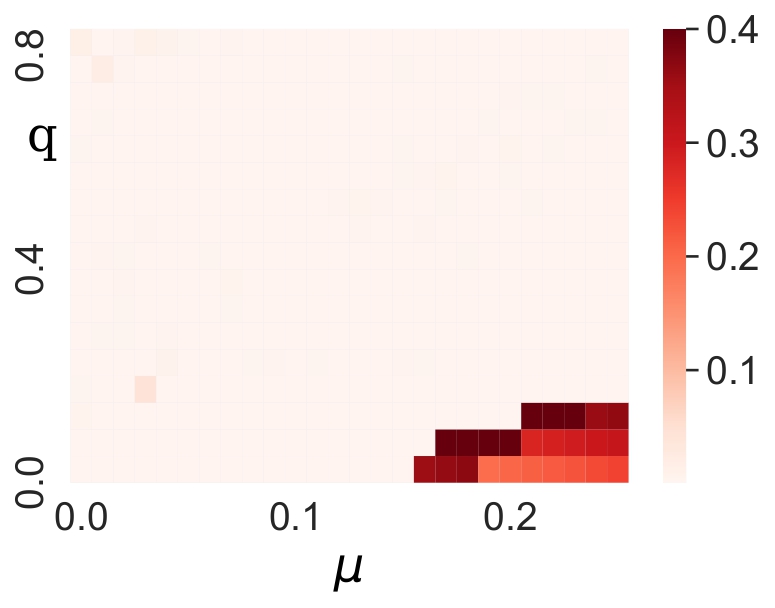}}
    \subfigure[]{\includegraphics[width=0.23\textwidth]{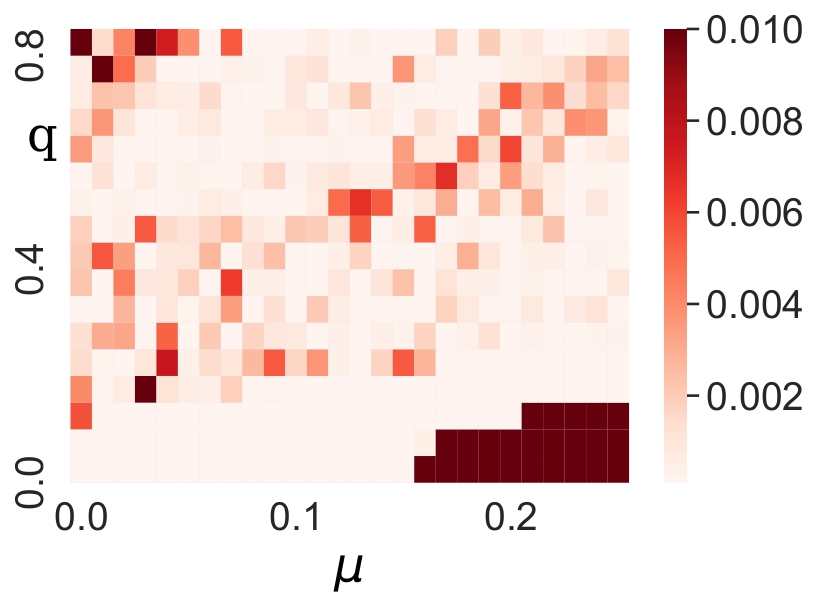}}
    \caption{\label{heatmap}In panel (a) density plots for the $\Delta$ (superconducting pair)  is shown. In panel (b) density plot of $m$ (magnetic order parameter) is presented. in $q-mu$ plane in the upper left and right panel respectively. In panel (c) density plot for  beam-splitter current is presented in $q-\mu$ planel. Finally in panel (d) zoomed in version of beam-splitter current is shown to illucidate the oscillating nature of current. }
\end{figure}

We numerically solve the self consistent equations for order parameters to get the corresponding values and use them to compute the current as given in Eq. \ref{fincurrent}. Before presenting our results for the beam-splitter current we briefly discuss the phase diagram in {\bf q} vs $\mu$ plane where {\bf q} is the nesting vector and $\mu$ denotes the amount of doping in the hole or electron pocket. In Fig. \ref{qmuphase} we present the phase diagram obtained after numerically solving the self-consistent equations as described in Sec.\ref{citeref}. The main feature of the phase diagram is that the co-existence phase which is the salient aspect of Iron-pnictide superconducting system exists for specific values of {\bf q} and $\mu$. For example, to get the co-existence phase a finite nesting(finite values of {\bf q}) is needed as evident from Fig. \ref{qmuphase} .  Also for very large values of {\bf q}, the coexistence phase ceases to exist yielding only superconducting phase.  Here we briefly compare our numerical phase diagram to that obtained in  reference \onlinecite{chubukov2}.  For zero nesting case, i.e. {\bf q}=0 we observe  a transition from pure SDW to pure SC phase around 0.15 doping  or the mismatch factor between the hole and electron pocket as previously obtained at zero temperature\cite{chubukov2}. It is important to note that $\mu$ is same as the $\delta_0$ considered there.  Also for {\bf q}$\neq 0$ case, the transition happens at larger values of $\mu$ than which is obtained at {\bf q}$=0$. Later we  present  further comparison of density of state calculation with the previous results in Sec. \ref{osc-dos} which has been used to benchmark our result. In our numerical analysis, we have taken $V_s = 2.5$ and $V_m = 3.0$ and chemical potential is  initially kept  at 1.7 which determines  the initial pocket sizes(i.e. before doping) to standardise our results with the previous work\cite{chubukov2}. Value of chemical potential is taken in the units of meV. It may be noted that in Eq. \ref{ekcf} $\boldsymbol{k}^2$ is to be understood as $\hbar^2 \boldsymbol{k}^2$.  We know that using value of $|\boldsymbol{k}| \sim 2 \pi/a$ where $a$ is of the order of Angstrom, one finds $\frac{\hbar^2 \boldsymbol{k}^2}{2m}$ of the order of meV. In our numerical calculation we have taken $m_c=m_f=1$ in Eq. \ref{ekcf}(instead of 0.51 meV) which  implies that the lattice spacing $a$ to be appropriately scaled to a higher value $a \rightarrow \sqrt{2.0}$ Angstrom (approximately) which is more realistic.  The value of $\bf q$ is to be understood of the order of $2 \pi/a$. Value of $\Delta$ is to be understood in meV as well. As we have plotted $\mathcal{I}_D$ instead of $\mathcal{I}_1$, the unit of $\mathcal{I}_D$ is $\rm{meV}^{-2}$ where as the unit of {\bf q} is $\rm{Angstrom}^{-1}$. Further we have confirmed the previous finding that the sign change of the superconducting gaps between the hole and electron pocket i.e $\Delta_c= - \Delta_f$ happens only when $V_s$ is positive. To complete the understanding of phase diagram given in Fig. \ref{qmuphase}, we have plotted the  amplitude of superconducting gap $\Delta$ and magnitude of magnetic order parameter $\bf m$ in {\bf q}$-\mu$ plane in Fig. \ref{heatmap} (a) and (b) respectively. These clearly shows that only in the coexistence phase both the parameters are finite. We refer reference \cite{chubukov2, araujo, bang, hu-ting-zhu, jiang-li-wang} for more details on the mechanism of magnetic and superconducting ordering and the phase transitions for the interested reader. 

Finally In Fig. \ref{heatmap} (c), we plotted the beam-splitter current  in the {\bf q}$-\mu$ plane for $k_f \delta r =0 $  which clearly shows that in the superconducting phase at low {\bf q} regime current is significantly higher than in the coexistence phase. However in the coexistence phase the current is finite though lower in magnitude  than that in pure superconducting phase. If we zoom in the current in coexistence phase as given in Fig. \ref{heatmap} (d), we find that the current is higher in some region and lower in some region depending on the values of {\bf q} and $\mu$. In subsequent discussion we describe in detail how the current behaves as a function of $k_f \delta r$ for  various fixed values of $\mu$ and {\bf q}.
\subsection{Comparison of beam-splitter current for electron doped vs hole doped}
\begin{figure}[h!]
\includegraphics[height=5cm,width=8cm]{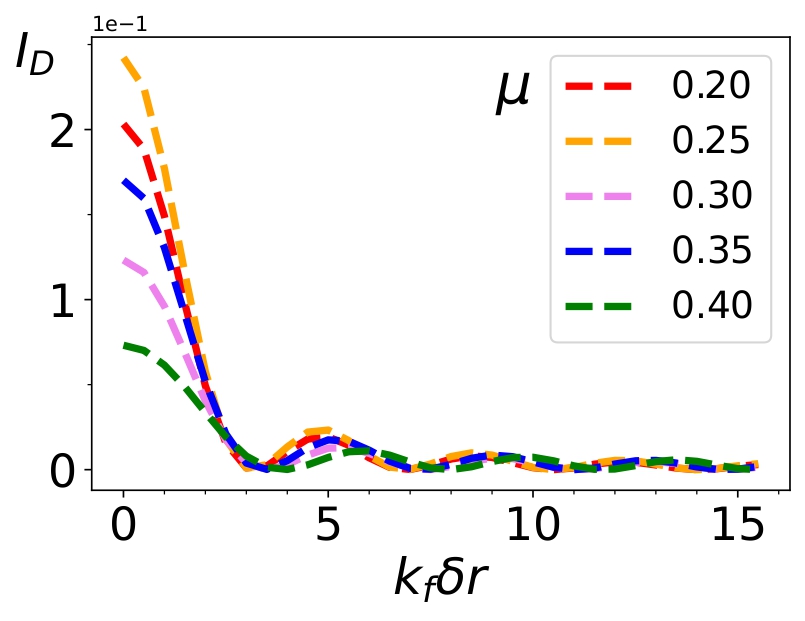} \\
\includegraphics[height=5cm,width=8cm]{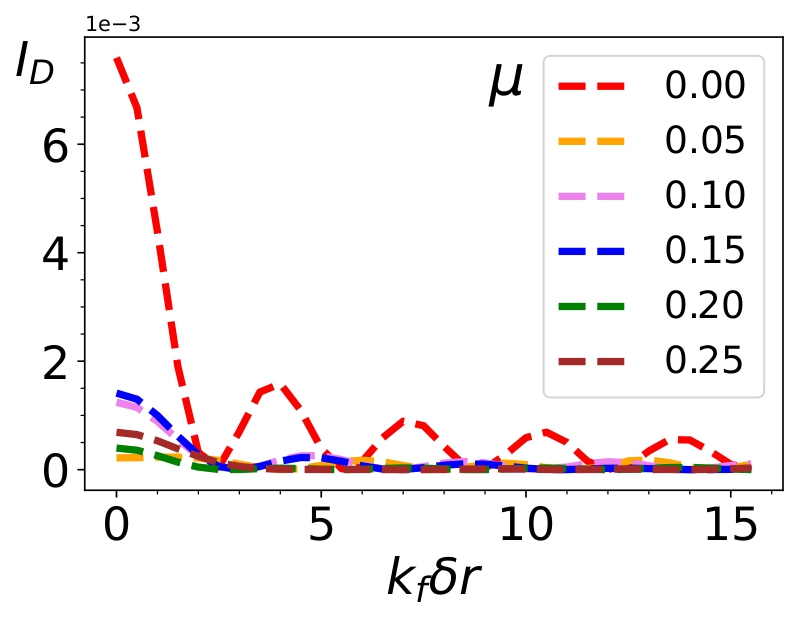}
\caption{\label{muceqmuf} Beam splitter current is plotted for identical shapes of electron and hole pocket for different chemical potential. Here the chemical potential $\mu$ refers  both the electron and hole doping. In the upper panel we plotted current for perfect nesting {\bf q}=0 and in the lower panel we have plotted current for finite nesting {\bf q}=0.4.}
\end{figure}
Here we discuss how the current varies with distance between the leads (which we denoted as $\delta r$) when the chemical potential at hole and electron pocket are varied independent of each other. First we discuss what happens when there is perfect nesting such that $\bf{q}=0$. In this case we note that there is no co-existence phase and magnetic order parameter does not exist. As a result there is no such entanglement between the hole and electron pocket though the superconducting gap in hole pocket is determined by the electrons at electron pocket through self-consistent equations. Throughout this article we have taken identical shape of the hole and electron pocket which is circular in our case, for simplicity.  In the upper panel of  Fig. \ref{muceqmuf} we plotted the current when the nesting vector $|q|$ is zero and in the lower panel we have plotted for a finite nesting vector $\bf{q}$. Due to symmetry the  beam current does not depend on the nature of doping whether it is electron doped or hole doped. The chemical potential $\mu$ in Fig. \ref{muceqmuf} represents $\mu_c$. The identical plots are obtained for $-\mu$ which represents electron doping i.e $\mu_f$.  This is consistent with the symmetry of the system.  In both the cases, however, the current is non-linear for a given value of $k_f \delta r$. For example when  $\mu$ is increased from $0.2$ to $0.25$ in the upper panel of Fig. \ref{muceqmuf}, we find an increase in current but further increase in $\mu$ to $0.4$ causes a decrease in current for perfect nesting. Similar observation holds for finite nesting as well. The main difference between the upper and lower panel of the Fig. \ref{muceqmuf} is that the magnitude of current is decreased by two decimal magnitude for finite $\bf{q}$. Noting that a finite $\bf{q}$ represents a co-existence phase having magnetic and superconducting order parameter both, it can be concluded that existence of magnetic order parameter is the main reason behind this decrease of current. To understand how this happens we note the form of $\mathcal{A}_k$ given in Eq. \ref{formA} can be written as $\mathcal{A}_{k,c}+ \mathcal{A}_{f,k}$ where $c,f$ denote the contribution from electron and hole pocket respectively. We notice that in pure superconducting phase the current due to hole and elctron pocket is additive and there is no cross terms. On the other hand in the presence of magnetic order the magnitude of $\mathcal{A}_{c,k}$ or $\mathcal{A}_{f,k}$ is decreased individually due to normalization factor now spread over both kind of fermions.  Also in the coexistance phase the cross terms in the expression $\mathcal{A}_k$ does not vanishes, i.e there is quantum entanglement between the hole current and electron current which results in quantum interference effect to set in.  These two facts make the current decreased in comparision to pure superconducting phase.
However the property which is common  in both the cases (finite and zero nesting) is that current is oscillatory or non-monotonous  at a given $k_f \delta{r}$ with respect to varying $\mu$.   While the reason of oscillations for zero nesting may be attributed to density of states variation with the system parameters, the oscillations at finite nesting comes from quantum interference effect due to entanglement between hole and electron pocket.

\subsection{Variations of beam splitter current with respect to nesting vector q}
\begin{figure}[h!]
\includegraphics[height=4.9cm,width=8cm]{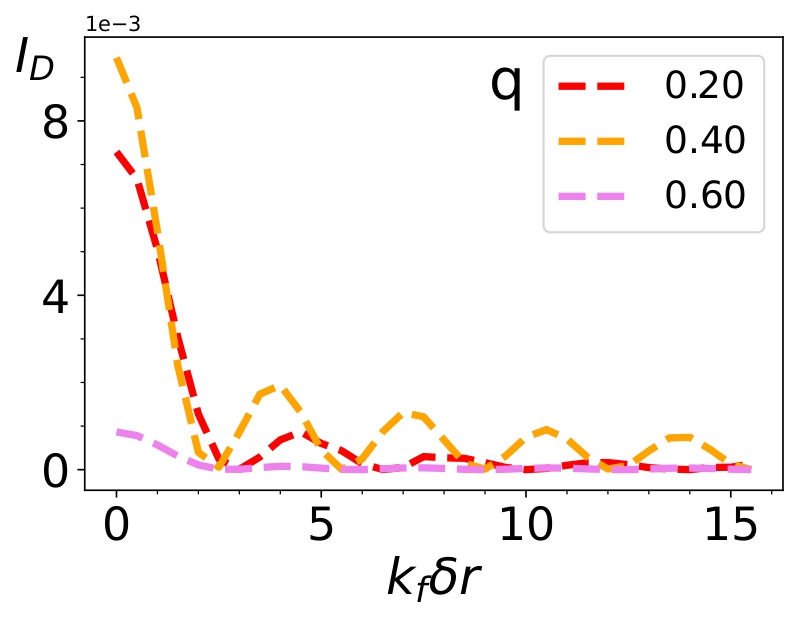}
\includegraphics[height=4.9cm,width=8cm]{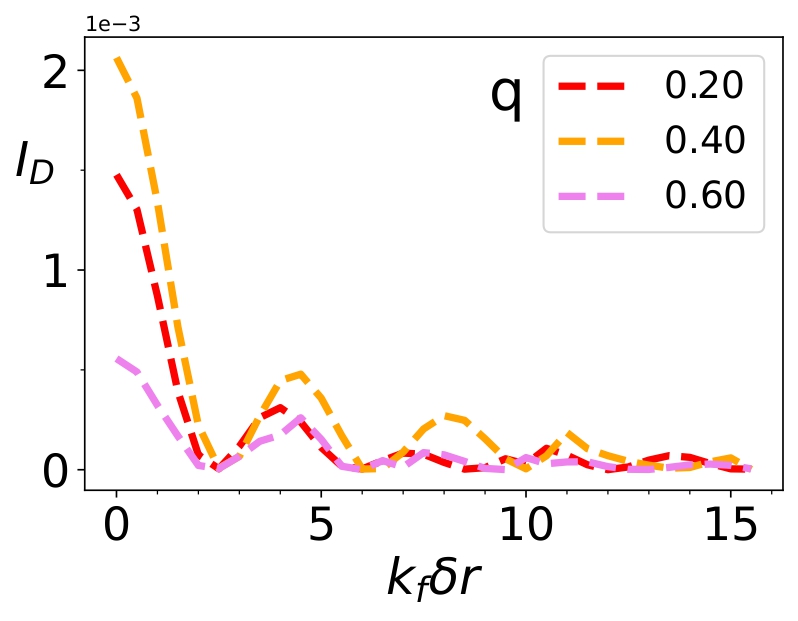}
\caption{\label{mucnemuf1}In the upper panel beam-splitter current is plotted for equal size hole and electron pocket (i.e $\mu=0$). In the lower panel current is plotted for electron pocket bigger than hole pocket i.e for case of electron doped system. We have taken $\mu=0.2$. The current is same if the hole pocket is doped by same amount. For both the cases, different colors represent  different amount of finite nesting for which currents have been plotted.}
\end{figure}
Having discussed the effect of doping at electron and hole pocket on beam current, we move on to discuss how the current behaves for non-zero nesting vector for different sizes of the hole and electron pocket. A non-zero nesting vector imply that if the electron pocket is situated at $(0,0)$, the
hole pocket is situated at $\bf{Q} + \bf{q}$ where $\bf{Q}=(\pi, \pi)$ and $\bf{q}$ is called nesting vector. We also note that for non-zero $\bf{q}$, magnetic ordering is present along with superconducting ordering. In Fig. \ref{mucnemuf1} upper panel we have taken electron and hole pocket to be of identical size i.e no doping and plotted current for representative value of  $\bf{q}$. As can be seen, in this case also, the current varies in a non-linear fashion with respect to $\bf{q}$ at given values of $k_f\delta r$. In the  lower panel we have plotted the current for hole doped system   for various values of $\bf{q}$ (for electron doped system, the results is identical). This implies a situation where  one pocket is bigger than the other. In this case also the beam splitter current is non-linear with respect to $\bf{q}$.  However for both the  cases we observe that as the magnitude of nesting vector increases, the beam current decreases and  there is an oscillation for intermediate values of nesting vector. This oscillation arises due to same reason explained in previous section.
\subsection{Oscillations of beam-splitter current}
\label{osc-dos}
\begin{figure*}[!tbp]
    \centering
    \includegraphics[height=10.9cm,width=18cm]{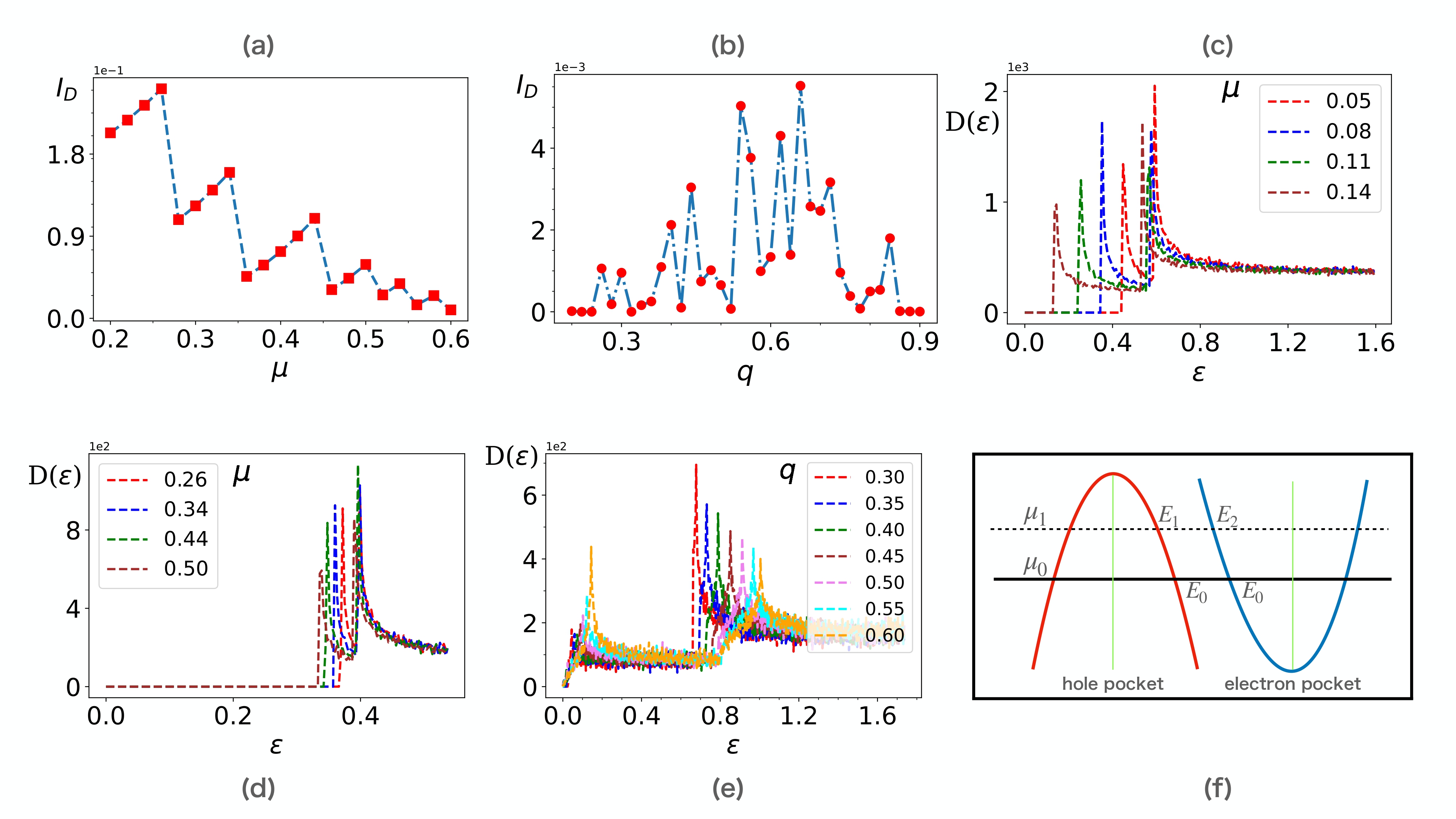}
    \caption{\label{oscmuq} In the panel (a) beam  splitter current is plotted for different dopings for {\bf q}=0 at  $k_f \delta r=0$. In the panel (b) beam splitter current is plotted against $q$ for $\mu=0$ and $k_f \delta r=0$ . In both the cases oscillations are observed. In the panel (c) and (d) we plot density of states for representative values of chemical potential ($\mu$) for $q=0$. Note that according to Fig. \ref{qmuphase} panel (c) corresponds to a purely magnetic phase. In the panel (e)  we plot density of states for various values of $q$ at $\mu=0.2$. In panel (f), the effect of electron doping is pictorially shown. }
\end{figure*}
In the previous two sections we have explained that there are characteristic oscillations of beam splitter current. For a given $k_f \delta r$ the current depends on the chemical potential or doping in a non-linear way. To show these oscillations in a more transparent way, in Fig. \ref{oscmuq}, we plot the beam splitter current for different doping as well as for different nesting vectors at  $k_f \delta_r =0$. In the  panel (a) we plot the current for a given value of $k_f \delta r$ for $q=0$. Evidently  current decreases with respect to $\mu$ as  expected but for intermediate ranges of $\mu$ current increases. This is the result of interdependence of superconducting pairing order ($\Delta_c$ and $\Delta_f$) on electron and hole pocket  as discussed before. We note that panel (a) in Fig. \ref{oscmuq} refers to pure superconducting phase and there is no magnetic ordering  present.  In this situation the current due to hole and electron pockets add up triviallly though there is indirect connection between the two current in the sense that the $\Delta_c$ is determined by the electron pocket ($f$-fermions) and $\Delta_f$ is determined by hole pocket ($c$-fermions). As chemical potential is increased, the effective momentum space area determining $\Delta_c$(or $\Delta_f$)  increases or decreases. This results in interdependency of density of states  on chemical potential. In the extreme limit when chemical potential is increased beyond a critical value, the superconductivity vanishes and current also vanishes as evident from panel (a) in Fig. \ref{oscmuq}. \\
\indent

In the panel (b) we plotted the  beam splitter current for various values of $\bf{q}$ for  $\mu=0.2$ and $k_f \delta r=0$. We see that depending on the values of $\bf{q}$, current  increases or decreases. We note that a finite $\bf{q}$ represents  co-existence phase with simultaneous presence of magnetic and superconducting orderings in the system. The nature of oscillation is different than the previous case. Here we find a Gaussian-like distribution pointing out that there is an optimal value of the nesting vector for which one obtains maximum beam splitter current.  To understand this oscillations we note that in the coexistence phase an eigenvector has the linear combinations of electrons and holes from both the pocket. This quantum entanglement resuts in specific interference pattern which yields this oscillations,  much similar to double-slit difraction pattern there is central peak with maximum height and subsequent reduced peaks at the two sides of it. \\
\indent

 We have shown the density of states  for different values of $\mu$ (for $q=0$)  in panel (c) and (d) and  for $q$ (and $\mu=0.2$) in panel (e) respectively. In the panel (c), the density of state corresponds to  pure magnetic phase while panel (d) corresponds to a pure superconducting phase. The panel (e) corresponds to a coexistence phase. For the pure magnetic phase as in panel (c) the density of states are expected to be peaked around $m +\mu$ for excitation at hole pocket and around $m-\mu$ for electron pocket where $m$ is the magnetic order parameter. This is due to the reason that the modified chemical potential after meanfield approximation are $m \pm \mu$ for electron and hole pocket respectively.  For $\mu=0.05$, our phase diagram in Fig. \ref{heatmap} yields $m \approx 0.6$ and we obtain the peak of density of states around $0.6$. As we increase the value of $\mu$, depending on the resulting magnetic order the two peaks occur at approximately $m \pm \mu$.  \\
\indent
For the pure superconducting phase  the density of states are plotted in panel (d) Fig. \ref{oscmuq}. We note that for a normal one pocket BCS superconductor the density of state is given by $D_s(E)= \frac{D_n(E_f) E}{\sqrt{E^2- \Delta^2}}$ where $\Delta$ is the superconducting gap and $D_n(E_f)$ is the normal state density of states at $E= E_f$. The density of state shows a divergence at $E= \Delta$ which is due to the sum rule $ D_s(E_s) dE_s=  D_n(E_n) E_n$ where $D_s$ and $D_n$ are the density of states in superconducting and normal state respectively. $E_s$ and $E_n$ are  corresponding energies in the two phases. One can check that for $\mu_c=\mu_f=0=m$, the eigenenergies of the two kind of excitations become identical (with $B_k=0, \Delta_c=-\Delta_f=0.4$) and density of states shows a divergence at $E=0.4$. In the presence of finite doping $\mu_{c/f} \ne 0$, this single peak seperates into two peaks as evident in panel D, Fig. \ref{oscmuq}.
\indent
For finite $\bf q$, one gets coexistence phase where superconductivity and  magnetic ordering  both exist. Such coexistence implies that the effective density of states contributing to superconductivity decreases as the sum rule $ D_s(E_s) dE_s=  D_n(E_n) E_n$  no longer holds. Hence we expect a decrease of current as evident from Fig. \ref{oscmuq}, the current is two order of magnitude less than the current found in only superconducting phase. The presence of magnetic ordering along with superconductivity causes broadening of the density of states peaks as shown in panel (e) in Fig. \ref{oscmuq}. In panel (e), Fig. \ref{oscmuq} we have taken $\mu=0.2$ and values of $\bf q$ varies from 0.3 to 0.6.  From panel (b), Fig. \ref{heatmap} we find that for these values of $\bf q$ values of $m$ ranges increases
from $0.2$ to $0.4$ and the magnitude  of $\Delta_c$ and $\Delta_f$  remain close to $0.4$. 
\section{Beam current through the same dots $\mathcal{I}_2$}
\label{i2}
 Till now we have discussed the scenario where only one of the two electrons(or holes) from the electron (or hole) pocket is collected at either dot and the second electron (or hole) is collected at remaining dot. The current generated from this process is the entangled beam-splitter current which is  matter of interest to us. However the alternative process such as tunneling of two electrons (or holes) via same dot is a real possibility which yields a contribution to current itself. There are two processes in which electron participating in the Cooper pair can tunnel via same dot. Case (I): First one electron(or hole) tunnels from the superconductor to dot-A let's say, and then following it the second electron(or hole) also tunnel to the same dot, as a result there are two electrons (or holes) at the same dot which costs an additional Coulomb repulsion energy of U.  Thus this virtual state is suppressed by a factor 1/U. Finally the two electrons(holes) leave the dot and tunnel to the lead.  Case (II): In the other process one electron (or hole) tunnels to say dot-A and then subsequently leaves the dot and tunnels further to lead-A, leaving an excitation on superconductor that costs additional  energy $\Delta$ before finally the second electron tunnels to the lead-A via dot-A.   The current taking into account these two processes  can be written as,  
\begin{eqnarray}
\mathcal{I}_2 && = \frac{2e\gamma_{s}^2\gamma}{\pi^2\nu_s^2}{\mathcal{I}_S},~~~~~ \mathcal{I}_S= (\sum_{k}{\mathcal{A}^{\prime}_k-\mathcal{A}^{\prime\prime}_k})^2
\label{isamedot}
\end{eqnarray}
$\gamma_s= 2 \pi \nu_s |T_{DS}|^2$ with $\nu_s$ is the density of states of the superconductor. $\gamma=\gamma_1+ \gamma_2$
with $\gamma_l= 2 \pi \nu_l |T_{DL}|^2$ with $\nu_l$ being the density of states at lead '$l$'. The expression for $\mathcal{A'}_k$ and $\mathcal{A}^{\prime\prime}_k$
are given below.
\begin{eqnarray}
&&\mathcal{A}^{\prime}_k  = \frac{u^c_{1,k}u^c_{2,k}}{E_{1,k}^2} + \frac{u^f_{1,k}u^f_{2,k}}{E_{1,k}^2}+\frac{u^c_{3,k}u^c_{4,k}}{E_{2,k}^2} + \frac{u^f_{3,k}u^f_{4,k}}{E_{2,k}^2} ~~~~~~~~~~~ \\
&&\mathcal{A}^{\prime\prime}_k  = \frac{u^c_{1,\tilde{k}}u^c_{2,\tilde{k}}}{E_{1,k}^2} + \frac{u^f_{1,\tilde{k}}u^f_{2,\tilde{k}}}{E_{1,k}^2} +\frac{u^f_{3,\tilde{k}}u^f_{4,\tilde{k}}}{E_{2,k}^2} + \frac{u^f_{3,\tilde{k}}u^f_{4,\tilde{k}}}{E_{2,k}^2} 
\end{eqnarray}
In the above $\tilde{k}=-k$ and expressions for $E_{i,k}, u^c_{i,k}, u^f_{i,k}$ and $u^c_{i,\tilde{k}}, u^f_{i,\tilde{k}}$ are given in the Appendix C. Now in the $\mathcal{I}_S$ as given in Eq. \ref{isamedot}, there are two contributions for above mentioned processes  which would add up. For detailed calculation of current we request the reader to look into the Appendix-A. To have a comparison between the current as given in Eq. \ref{fincurrent} and in Eq. \ref{isamedot}, we estimate the ratio between the two. It is straightforward to obtain, 
\begin{eqnarray}
\frac{\mathcal{I}_1}{\mathcal{I}_2} = \frac{\mathcal{I}_D}{2\gamma^2{\mathcal{I}_S}}
\end{eqnarray}
\begin{figure}[h!]
\includegraphics[height=5.5cm,width=8.1cm]{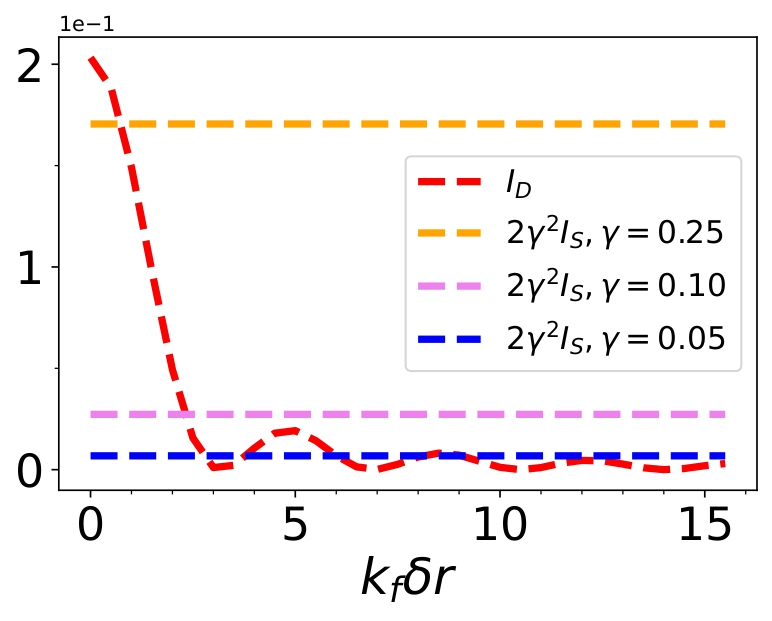}
\caption{\label{suppress}A comparison of $\mathcal{I}_D$ and $\mathcal{I}_S$. The oscillating current represents $\mathcal{I}_D$ for $\mu=0.20$. The constant currents correspond $I_S$ for various values of $\gamma$. }
\end{figure}
Clearly $\mathcal{I}_D>2\gamma^2{\mathcal{I}_{S}}$ is the the desired regime where $\mathcal{I}_1$ dominates $\mathcal{I}_2$. As shown in Fig. \ref{suppress}, for $\gamma=0.25$ $\mathcal{I}_1$ is larger than $\mathcal{I}_2$ for some finite values of $k_f \delta r$. We also observe that a decrease of $\gamma$ from $0.25$ to $0.10$ rapidly increases the efficiency of beam-splitter current to larger values of $k_f \delta r$. \\
\indent

Before we conclude and summarize our study, it is imperative to compare the beam-splitter arrangement studied in earlier pioneering work by Daniel loss and collaborators \cite{daniel} and subsequent extension  for graphene based superconductor by Firoz Islam et.al \cite{firoz}. Apart from the inherent entangled nature of beam-splitter current, the salient feature of the beam-splitter current for an ideal s-wave BCS supercurrent is characteristic oscillation of current with $k_f \delta r$ which is also present in graphene based superconductor and also in the present study. Whereas the chemical potential determines the overall magnitude of the beam-splitter current. However there is an important difference found on the dependency of magnitude of beam-splitter current on  chemical potential for normal BCS superconductor (and graphene based superconductor) with the iron-pnictide superconductor considered here. The beam-splitter current monotonically decreases as the chemical potential is decreased for ordinary BCS superconductor and graphene based superconductor. But for iron-pnictide system the dependence of overall magnitude of current is non-monotonous with the chemical potential. The reason is the quasi-particle contributing to the beamsplitter current originates from a single fermi-pocket in case of BCS superconductor and graphene based superconductor and thus changing chemical potential has monotonous dependency on the current. However for iron-pnictide the contribution to the beam-splitter current comes from quasi-particles originating from hole and electron pockets separeted by a momenta $Q$. The change of chemical potential effects the size of the electron and hole pocket in opposite way, i.e if electron pocket size is decreased, hole pocket size is increased. Due to the interdependency of beam-splitter current on both the pockets non-trivially, the current varies non-monotonously with the change on chemical potential.  Apart from this important difference there are some other significant differences between   iron-pnictide superconductor and  graphene based superconductor. First of all in graphene the mechanism of superconductivity is not yet  established and for this reason proximity based superconductor is assumed. The proximity induced supercondctor considered earlier included both s-wave and p-wave superconductor ordering defined as $\Delta_0= <\alpha_{i \uparrow} \alpha_{i \downarrow}>= <\beta_{i \uparrow} \beta_{i \downarrow}> $ and $\Delta_1 \sim <\alpha_{i\uparrow} \beta_{j\downarrow}-\alpha_{i\downarrow} \beta_{j\uparrow}> $. Here $\alpha$ and $\beta$ denote the two sublattices in honeycomb lattice. The problem finally reduces as two copies of BCS Hamiltonian interms of $c$ and $d$ operator which are linear combinations of original $\alpha$ and $\beta$ by $\alpha_{ks}=\frac{1}{\sqrt{2}}(c_{ks} + d_{ks}), \beta_{ks}=\frac{1}{\sqrt{2}} e^{-i\phi_k}(c_{ks}-d_{ks})$. The bogoliubov quasi-particle energy in the``$c$'' and ``$d$'' channels are given by  $E_{k,\nu}=\sqrt{(\xi_k+ \nu \mu)^2 + (g_0 \Delta + \nu g_1 \Delta_1 |\gamma_k|)^2}$, where $\nu=\pm$, where  $\nu=+$ for $c$-channel and $\nu=-$ for $d$-channel. The Bogoliubov quasi-particles in the $c$-channel are given by $c_{k,\uparrow}= u_{k,+}\gamma_{1k\uparrow} + v_{k,+} \gamma^{\dagger}_{1-k\downarrow}, c_{-k,\downarrow}= u_{k,+}\gamma_{1-k\downarrow} - v_{k,+} \gamma^{\dagger}_{1k\uparrow}$. The equivalent relations for the $d$-channels are given by $d_{k,\uparrow}= u_{k,-}\gamma_{2k\uparrow} + v_{k,-} \gamma^{\dagger}_{2-k\downarrow}, d_{-k,\downarrow}= u_{k,+}\gamma_{2-k\downarrow} - v_{k-} \gamma^{\dagger}_{2k\uparrow}$. Here $u_{k,\nu}= \frac{1}{\sqrt{2}}(1+ \frac{\xi_k}{E_{k,\nu}})^{1/2}$ and $v_{k,\nu}=\frac{1}{\sqrt{2}}(1-\frac{\xi_k}{E_{k,\nu}})^{1/2}$. It has been reported that when only $\Delta_0$ is present the beam-splitter current is larger than the case where $\Delta_0$ and $\Delta_1$ are both present. Though  for both the cases   the physical problem remains two separeted copies of BCS Hamiltonian, the total superconducting quasi-particle density changes when the combinations of order parameters are different. This is evident as the superconducting number density depends on $\sum_{k, \nu} u_{k\nu} v_{k\nu}$ and $u_{k\nu}$ and $v_{k\nu}$
which depends on $E_{k,\nu}$ through $\Delta_0$ and $\Delta_1$ differently. Thus the physics relies on the re-distribution of quasi-particle density in each blocks. The relation between $c$ and $d$ fermion on the Bogoliubov quasiparticles can be readily compared with the Eqn \ref{bogo1} to Eqn \ref{bogo4}. It is evident
that in the present case the Bogolubov quasi-particles in $c$ and $f$ channels are not decoupled due to the
presence of magnetic order parameters. In the  graphene based superconductor such couplings  are absent. This coupling between the hole and electron pocket  which gives rise to magnetic order parameters are the crucial ingredients for the iron-pnictide based beam-splitter studied here. This gives rise to the non-trivial interference pattern between the two bogoliubov quasiparticles coming from hole and electron pockets.

\section{Discussion}
\label{discussion}
To summarize we have proposed a Cooper pair beam splitter arrangement for Iron-pnictide superconductor which is known for hosting Cooper pairs responsible for superconducting property as well as inherent magnetic ordering. The co-existence of these two seemingly non-inclusive order  parameters in a given material provides  an interesting  platform to investigate the consequences in a Cooper pair beam splitter current. We have considered various realistic situations found in actual materials such as zero nesting and finite nesting \cite{chubukov1}, equal and unequal size of hole and electron pockets etc. Most notably our finding indicates that in general  the beam splitter current depends  non-monotonically on electron $\mu_c$ and hole doping $\mu_f$ as well as on the magnitude of nesting vector $|{\bf q}|$. In all these cases there are a critical values for which the current is maximum. We believe that this fact might be useful in practical application such as switching and quantum gate applications. In the absence of magnetic ordering the beam splitter current is additive due to hole and electron pocket.  However in the co-existence phase due to the entanglement between fermions in electron and hole pocket  there are quantum  interference effect which results in characteristic oscillations of beam splitter current with respect to nesting vector. \\
It may also be noted that though in the beamsplitter set up we have assumed $|T_{DS}|< |T_{DL}|$ we do not expect any Kondo effect to set in within the 
quantum dots because initially in the static limit the quantum dots are taken as vacuum state. Also the tunneling of electron from the lead to quantum dots are surpressed by the grand canonical distribution function $\infty ~\rm{exp}(-\Delta\mu/K_BT)$ where $\Delta\mu$ is the bias voltage. Even a negligible amount of tunneling of normal electron from lead to quantum dot are going to be in even numbers due to participation of electrons of either spins. This would also make the total number of electrons on quantum dots to be even number which works against Kondo effect to be realized. \\
 The effect of external pressure may be an useful way to control the nesting vectors as found before \cite{Vladimir}. It may be noted that we have done our calculation of beam splitter  current at zero temperature for a prototype iron-pnictide superconductor whereas in real material such superconductivity having coexistence of magnetic order  and superconducting order happens upto quite large temperature in comparison to conventional BCS superconductor. We leave the question of finite temperature effect of this beam splitter set up for a future study.   However given the fact that all the existing studies so far on beam splitter current is at zero temperature BCS like superconductor and the finite temperature stability of Iron-pnictide superconductor for a wide range of temperature, we think that the study done here would  be a realizable at higher temperature than a conventional BCS supercondctor based beam-splitter set up. \\ 

However we note that the magnitude of the Iron-pnictide based beam-splitter current is smaller in the co-existence phase  in order of magnitude than the conventional two dimensional BCS superconductor based  beam-splitter  current. With the recently proposed protocol of measuring current produced by single electron detection \cite{bylander,fujisawa} one can expect that the current may be measured.  However one also needs to consider the signal to noise ratio for such measurement. We think that our present study would motivate for further investigations on other aspects such as effect of ferromagnetic proximity effect \cite{hofstetter1,burset-2011}, effect of impurity \cite{hofstetter1,burset-2011} and Bells inequality\cite{samuelsson-2003}. \\

Finally we note that with  the recent significant developments in the field of  theoretical and experimental
topological insulator based superconductor named topological superconductor, the beam splitter arrangement
could serve as an unique set up to detect various quantum mechanical aspect of the topological states as well
as the superconducting properties. Recently superconductivity at room temperature has been observed in few-layer
Stanene which also host a quantum spin hall phase\cite{menghan-liao}. How the robustness of the topological states and superconducting state responds due to various perturbations and disorder is an interesting aspect that can be looked at these topological superconductors \cite{menghan-liao,shu-ping-lee,sean-hart,zhao}. Further we think the presence of inter-valley scattering in Chern insulator based moire superconductor or in twisted bilayer may have characteristic oscillations or interference pattern in the beam-splitter set up arrangement considered here \cite{chuan-wu,ashvin,efetov,ipsita}. However the scope of this study remains open in future.

\begin{acknowledgments}
SM thanks Arijit Saha for useful discussions and  S P Mandal for hospitality.
\end{acknowledgments}

\appendix
\section{Intermediate steps for current calculation}

\subsection{Intermediate steps for current $\mathcal{I}_1$}
Here we give detailed steps to arrive at the current expressions given in Eq. \ref{fincurrent} and Eq. \ref{isamedot}. We first  provide detail steps used to evaluate $\mathcal{I}_1$, the current when holes (or electrons) reaches different dots. The main step is to  evaluate the  transition amplitude $\langle f|T|i \rangle$ defined in Sec. \ref{sec:level2}.  We use Eq. \ref{totalT}, Eq. \ref{firstT} and Eq. \ref{secondT} to arrive at the following expression,
\begin{eqnarray}
\label{eqa1}
\langle f|T|i \rangle &&= \langle f|H_{DL}\sum_{n=0}^{\infty}\Big(\frac{1}{i\eta-H}H_{DL}\Big)^{2n+1} |f' \rangle\nonumber\\
&& \times \langle f'|\frac{1}{i\eta-H_{0}}H_{SD}\frac{1}{i\eta-H_{0}}H_{SD}|i \rangle \nonumber \\
&& = A_{D1}\times{A_{D2}} 
\end{eqnarray}

Where $A_{D2}$ and $A_{D1}$ represent the transition amplitude from supercondutor to quantum dots and from quantum dots to lead respectively and their expressions are as follows,

\begin{eqnarray}
\label{ad10}
A_{D1}&&=\langle f|H_{DL}\sum_{n=0}^{\infty}\Big(\frac{1}{i\eta-H}H_{DL}\Big)^{2n+1} |f' \rangle \\
\label{ad20}
A_{D2}&&=\langle f'|\frac{1}{i\eta-H_{0}}H_{SD}\frac{1}{i\eta-H_{0}}H_{SD}|i \rangle
\end{eqnarray}
Where $|f\rangle $ and $|f'\rangle $ are given in Eq. \ref{finstate} and Eq. \ref{finstate1}.  Here we note that $|i\rangle$ denotes the combined vacuum state for $\gamma_c, \gamma_f$ quasi-particles in superconductor, $d$-electrons at  dots and the  $a$-electrons at lead.
The  evaluation of $A_{D1}$ and $A_{D2}$ is given in detail below. For simplicity, in the below, we use $T^{\prime}$ and $T^{\prime\prime}$ (as defined in Eqs. \ref{firstT},\ref{secondT} ) and proceed. Then Eq. \ref{ad1} yields, 
\begin{eqnarray}
<f|T'|f'>&=&\frac{1}{2}<i|(a_{2q\downarrow}a_{1p\uparrow}-a_{2q\uparrow}a_{1p\downarrow})T'\times \nonumber\\
&&(d_{1\uparrow}^{+}d_{2\downarrow}^{+}-d_{1\downarrow}^{+}d_{2\uparrow}^{+})|i> \nonumber\\
&=&\frac{1}{2}(<i|a_{2q\downarrow}a_{1p\uparrow}T'd_{1\uparrow}^{+}d_{2\downarrow}^{+}|i> \nonumber\\
&&-\underbrace{<i|a_{2q\downarrow}a_{1p\uparrow}T'd_{1\downarrow}^{+}d_{2\uparrow}^{+}|i>} \nonumber\\
&& -\underbrace{<i|a_{2q\uparrow}a_{1p\downarrow}T'd_{1\uparrow}^{+}d_{2\downarrow}^{+}|i>} \nonumber\\
&&+<i|a_{2q\uparrow}a_{1p\downarrow}T'd_{1\downarrow}^{+}d_{2\uparrow}^{+}|i>)  
\end{eqnarray}
We note that in the above $a$-operators denote the fermions at the leads and $d$-operators denote the fermions at the quantum dots and we have used the spin-singlet
states at the lead and dots. Now the underbraced terms vanish because spin flip is happening which is not allowed by the design mechanism of the splitter.
Also it is interesting to notice that first and last terms  are identical under exchange of spin indices (${\uparrow\rightleftharpoons\downarrow} \nonumber$). All these imply that,
\begin{equation}
\label{ftprf}
     <f|T'|f'>=\frac{1}{2}2<i|a_{2q\downarrow}a_{1p\uparrow}T'd_{1\uparrow}^{+}d_{2\downarrow}^{+}|i> 
\end{equation}

Similarly for the Andreev process which denotes the tunneling of electrons from superconductor to quantum dots as denoted by Eq. \ref{ad20},
\begin{equation}
    <f'|T''|i>=\frac{1}{\sqrt{2}}<i|(d_{2\downarrow}d_{1\uparrow}-d_{2\uparrow}d_{1\downarrow})T''|i>
\end{equation}
Now it may happen that in the process of transport one electron with a particular spin(say up) comes to the dot from the superconductor but an electron with opposite spin(say down) may travel forward from the same dot to the lead, but we don't want that. We want processes like  $ |SS>\to|DS>\to|DD> $ and $|SS>\to|SD>\to|DD> $ such that entangled pair of electrons from the superconductor(SS) get transported to dot(DD) and move forward to separate leads, because we avoid spin flipping. Where, $|SD>=\sum_{\phi}\gamma^{\phi{+}}_{k\sigma}d_{l-\sigma}^{+}|i>$.
$\phi$ can take $c$ and $f$ two values due to two species of quasiparticles and $\gamma$ denotes the Bogoliubov quasiparticles in the superconductor. To proceed further we note  few important points for the below calculations; (a) $H_{S1D1}$ selects one electron of the entangled pair to dot 1 and $H_{S2D2}$ selects the other electron of the entangled pair to dot 2. (b) We take $+k$ for $\uparrow$ spin and $-k$ for $\downarrow$ spin without loss of generality. We also note that, in the following, we use the Bogoliubov transformations given in Eqs. \ref{bogo1},\ref{bogo2},\ref{bogo3},\ref{bogo4} and the Fourier transformations, $ \Psi^{c}_
    {\sigma}({\overline r_{l}})=\sum_{k}e^{i\overline k.\overline r_{l}}{c_{k\sigma}}=\sum_{-k}e^{-i\overline k.\overline r_{l}}{c_{-k\sigma}}$ and $ \Psi^{f}_
    {\sigma}({\overline r_{l}})=\sum_{k}e^{i\overline k.\overline r_{l}}{f_{k\sigma}}=\sum_{-k}e^{-i\overline k.\overline r_{l}}{f_{-k\sigma}} $ which are required while using $H_{SD}$
as given in Eq. 11. It is also  to be noted that since we work close to resonance we use $ |i\eta-E_{k}-\epsilon_{l}|\approx|E_{k}|$ where $E_k$ and $\epsilon_l$ denotes the energy of the Bogoliubov quasiparticle and energy of quantum dot. Now we proceed to simplify $<f'|T''|i>$ as follows,
\begin{widetext}
\begin{eqnarray}
\label{fti}
       <f'|T''|i>&=&\frac{1}{\sqrt{2}}<i|(d_{2\downarrow}d_{1\uparrow}-d_{2\uparrow}d_{1\downarrow}) \frac{1}{(i\eta-H_{0})}H_{s_{1}D_{1}}\sum_{k,l,\sigma,\phi=c,f}\gamma_{k\sigma}^{\phi+}d_{l-\sigma}^{+}|i><i|d_{l-\sigma}\gamma_{k\sigma}^{\phi}\frac{1}{(i\eta-H_{0})}H_{S_{2}D_{2}}|i> \nonumber\\
&=&\frac{1}{\sqrt{2}}<i|(d_{2\downarrow}d_{1\uparrow}-d_{2\uparrow}d_{1\downarrow})\frac{1}{(i\eta-H_{0})}[\sum_{l\sigma,\phi=c,f}T_{SD}d_{1\sigma}^{+}\psi^{\phi}_{\sigma}(\overline{r_{l}})+h.c]\sum_{k,l,\sigma,\phi=c,f}\gamma^{\phi{+}}_{k\sigma}d_{l-\sigma}^{+}|i>\nonumber\\
&&<i|d_{l-\sigma}\gamma^{\phi}_{k\sigma}\frac{1}{(i\eta-H_{0})}[\sum_{l\sigma,\phi=c,f}T_{SD}d_{2\sigma}^{+}\psi^{\phi}_{\sigma}(\overline{r_{l}})+h.c]|i> \nonumber\\
&=&\frac{1}{\sqrt{2}}[\sum_{k}<i|d_{2\downarrow}d_{1\uparrow}\frac{1}{(i\eta-H_{0})}T_{SD}d_{1\uparrow}^{+}[c_{k\uparrow}+f_{k\uparrow}][\gamma^{c{+}}_{k\uparrow}+\gamma^{f{+}}_{k\uparrow}]d_{2\downarrow}^{+}e^{i\overline{k}.\overline{r_{1}}}|i>\times \nonumber\\
&&<i|d_{2\downarrow}[\gamma^{c}_{k\uparrow}+\gamma^{f}_{k\uparrow}]\frac{1}{(i\eta-H_{0})}T_{SD}d_{2\downarrow}^{+}[c_{-k\downarrow}+f_{-k\downarrow}]e^{-i\overline{k}.\overline{r_{2}}}|i>] \nonumber\\
&&-\frac{1}{\sqrt{2}}[\sum_{k}<i|d_{2\uparrow}d_{1\downarrow}\frac{1}{(i\eta-H_{0})}T_{SD}d_{1\downarrow}^{+}[c_{-k\downarrow}+f_{-k\downarrow}][\gamma^{c+}_{-k\downarrow}+\gamma^{f+}_{-k\downarrow}]d_{2\uparrow}^{+}e^{-i\overline{k}.\overline{r_{1}}}|i>\times \nonumber\\
&&<i|d_{2\uparrow}[\gamma^{c}_{-k\downarrow}+\gamma^{f}_{-k\downarrow}]\frac{1}{(i\eta-H_{0})}T_{SD}d_{2\uparrow}^{+}[c_{k\uparrow}+f_{k\uparrow}]e^{i\overline{k}.\overline{r_{2}}}|i>] \nonumber\\
&=&\frac{1}{\sqrt{2}}[\sum_{k}\frac{{T_{SD}}^{2}}{\epsilon_{1}+\epsilon_{2}-i\eta}\mathcal{A_K}(\cos{\overline{k}.\delta\overline{r}})] \nonumber\\
\end{eqnarray}
\end{widetext}
Note that the $\sin({\overline{k}.\delta\overline{r}})$ term associated with imaginary part will go to zero while doing the $\theta$ integration in $k$ sum. The factor $\mathcal{A}_k$ is given in Eq. \ref{formA} and $\epsilon_1, \epsilon_2$ denote the  energies at quantum dot 1 and 2 respectively. We have mentioned earlier that there can be two ways of tunneling from superconductor to quantum dots. Both the processes  gives identical expressions which results in a multiplicative factor of 2. Incorporating this finally we obtain,
\begin{equation}
    <f'|T''|i>=\frac{1}{\sqrt{2}}[\sum_{k}\frac{{2T_{SD}}^{2}}{\epsilon_{1}+\epsilon_{2}-i\eta}\mathcal{A_K}(\cos{\overline{k}.\delta\overline{r}})] 
\end{equation}
Now lets calculate the second term as given in Eq. \ref{ftprf} in the below.
\begin{equation}
    <f|T'|f'>=\frac{1}{2}2<i|a_{2q\downarrow}a_{1p\uparrow}T'd_{1\uparrow}^{+}d_{2\downarrow}^{+}|i>
\end{equation}
The above process signify tunneling from quantum dots to the leads which are symbolically denoted as $|DD>=|LD>=|DD> $ and $ |DD>=|DL>=|DD>$. Here $|LD>$ (or$|DL>$ )
denotes the state after one electron reaches to first (or second) lead via respective quantum dots. In the subsequent steps we  denote the lead states $|LL>$ as $|pq>$ for notational conveniences. From the detail expressions of $ <f|T'|f'> $ from Eq. \ref{ad1} we note that various intermediate steps are to be inserted appropriately  as allowed by Wick's theorem and orthogonality condition. Considering all such possible processes we can write down,
\begin{eqnarray}
\label{pqdd}
<f|T'|f'>&=&<pq|T'|DD> \nonumber\\
&=&[<pq|H_{D_{1}L_{1}}|Dq> \nonumber\\
&&\times<Dq|(\sum_{n=0}^{\infty}(\frac{1}{i\eta-H_{0}}H_{D_{1}L_{1}})^{2n}|Dq> \nonumber\\
&&\times<Dq|\frac{1}{i\eta-H_{0}}H_{D_{2}L_{2}}|DD>+\nonumber\\
&&<pq|H_{D_{2}L_{2}}|pD> \nonumber\\
&&\times <pD|\sum_{n=0}^{\infty}(\frac{1}{i\eta-H_{0}}H_{D_{2}L_{2}})^{2n}|pD> \nonumber\\
&&\times<pD|\frac{1}{i\eta-H_{0}}H_{D_{1}L_{1}}|DD>] \nonumber\\
&&\times <DD|\sum_{m=0}^{\infty}(\frac{1}{i\eta-H_{0}}H_{DL})^{2m}|DD> \nonumber\\
\end{eqnarray}
Just calculating like before it is straightforward to have \cite{daniel},
\begin{eqnarray}
&& <DD|\sum^{\infty}_{m=0}(\frac{1}{i\eta-H_{0}}H_{DL})^{2m}|DD>= \\ \nonumber
&&\frac{1}{1- <DD|(\frac{1}{i\eta-H_{0}}H_{DL})^{2}|DD> }
\end{eqnarray}
where $ <DD|(\frac{1}{i\eta-H_{0}}H_{DL})^{2}|DD>= \frac{\Sigma}{i \eta-\epsilon_1-\epsilon_2}, \Sigma=|T_{DL}|^2 \sum_{lk}(i\eta-\epsilon_l-\epsilon_k)^{-1}$. In the presence of a Fermi sea in the leads we introduce a cutoff in the sum in $\Sigma$ at the fermi level given by $\epsilon_k \sim - \Delta \mu$ and at the edge of the conduction band given by $\epsilon_c$. Then one obtains $\Sigma=  \gamma_l {\rm ln}(\epsilon_c/\Delta \mu) - i \gamma/2$. Now we note that logarithmic renormalization of the self energy is small i.e., $\sim \gamma_l {\rm ln}(\epsilon_c/\Delta \mu) < \Delta \mu$ and for this we neglect this in the following. Here we have used $ \gamma_{l}=2\pi\nu_{l}|T_{DL}|^2, \gamma= \gamma_1 + \gamma_2 $, where $\nu_l$ is the DOS at the chemical potential $\mu_l$. All these yields the following simplification,

\begin{equation}
\label{dddd}
<DD|\sum^{\infty}_{m=0}(\frac{1}{i\eta-H_{0}}H_{DL})^{2m}|DD>= \frac{\epsilon_1 + \epsilon_2 - i \eta}{\epsilon_1 + \epsilon_2 - i \gamma/2}
\end{equation}

Similarly,
\begin{equation}
\label{pddd}
<pD|\sum^{\infty}_{n=0}(\frac{1}{i\eta-H_{0}}H_{D_2L_2})^{2n}|pD>= \frac{\epsilon_p + \epsilon_2 - i \eta}{\epsilon_p + \epsilon_2 - i \gamma_2/2}
\end{equation}

and 

\begin{equation}
\label{dqdd}
<Dq|\sum^{\infty}_{n=0}(\frac{1}{i\eta-H_{0}}H_{D_1L_1})^{2n}|Dq>= \frac{\epsilon_q + \epsilon_1 - i \eta}{\epsilon_q + \epsilon_1 - i \gamma_2/2}
\end{equation}
 Inserting Eqs. \ref{dddd}, \ref{pddd} and \ref{dqdd} in Eq. \ref{pqdd} we obtain,
\begin{eqnarray}
      <pq|T'|DD>&=&[\frac{|T_{DL}|^2}{(i\eta-\epsilon_{1}-\epsilon_{q})}\times\frac{(\epsilon_{1}+\epsilon_{q}-i\eta)}{(\epsilon_{1}+\epsilon_{q}-i\frac{\gamma_{1}}{2})}] \nonumber\\
      &&+\frac{|T_{DL}|^2}{(i\eta-\epsilon_{2}-\epsilon_{p})}\times\frac{(\epsilon_{2}+\epsilon_{p}-i\eta)}{(\epsilon_{2}+\epsilon_{p}-i\frac{\gamma_{2}}{2})}] \nonumber\\
     && \times\frac{(\epsilon_{1}+\epsilon_{2}-i\eta)}{(\epsilon_{1}+\epsilon_{2}-i\frac{\gamma)}{2}} \nonumber\\
      &&=-\frac{|T_{DL}|^2(\epsilon_{1}+\epsilon_{2}-i\eta)}{(\epsilon_{1}+\epsilon_{q}-i\frac{\gamma_{1}}{2})(\epsilon_{2}+\epsilon_{p}-i\frac{\gamma_{2}}{2})}\nonumber\\
\end{eqnarray}

Thus the final expression for $AD_1$ and  $AD_2$ as obtained are given below,
\begin{eqnarray}
\label{ad1}
A_{D1} && = -\frac{|T_{DL}|^2(\epsilon_1+\epsilon_2-i\eta)}{(\epsilon_1+\epsilon_q-i\gamma_1/2)(\epsilon_2+\epsilon_p-i\gamma_2/2)} \\
\label{ad2}
A_{D2} && = \frac{1}{\sqrt{2}}\Big[\sum_{k}\frac{{2 |T_{SD}|^2}}{\epsilon_{1}+\epsilon_{2}-i\eta}\mathcal{A}_k(\cos{\overline{k}.\delta\overline{r}})\Big]
\end{eqnarray}

In the above $\epsilon_1$ and $\epsilon_2$ represent the energy of quantum  dot 1 and dot 2 respectively.  The factor containing $\eta$ cancells due to contribution from $A_{D1}$ and $A_{D2}$. Also we note that in obtaining $AD_1$, we have removed the $\eta$  factors in the denominators as they are replaced by $\gamma_l$ at the end.  Note that the energy of the Bogoliubov quasiparticle $\mathcal{E}_{kq\pm}$ does not appear in the above two expressions as the initial state $|i \rangle$ is taken as vacuum of Bogoliubov quasiparticle. The full expression of $\mathcal{A}_k$ (as obtained in \ref{fti}) is given below.

\begin{eqnarray}
\mathcal{A}_k&&= \frac{u^c_{1,k}u^c_{2,k}- u^c_{1,-k}u^c_{2,-k}}{\mathcal{E}_{kq+}} + \frac{u^f_{1,k}u^f_{2,k}- u^f_{1,-k}u^f_{2,-k}}{\mathcal{E}_{kq+}} \nonumber \\
+ && \frac{u^c_{3,k}u^c_{4,k}- u^c_{3,-k}u^c_{4,-k}}{\mathcal{E}_{kq-}} + \frac{u^f_{3,k}u^f_{4,k}- u^f_{3,-k}u^f_{4,-k}}{\mathcal{E}_{kq-}}
\end{eqnarray}
Using Eq. \ref{ad1} and Eq. \ref{ad2} we arrive at the current expression given in Eq. \ref{fincurrent}.

\subsection{Intermediate steps for current $\mathcal{I}_2$}
Now we outline the intermediate steps to arrive at the expression for $\mathcal{I}_2$. Similar to Eq. \ref{eqa1} we find the following expression for the transition amplitude $\langle f| T| i \rangle$,  
\begin{eqnarray}
\langle f|T|i\rangle &&= \sum_{p''\sigma}\langle f|H_{DL}\sum_{n=0}^\infty\Big(\frac{1}{i\eta-H_{0}}H_{DL}\Big)^{2n}|f' \rangle\times\nonumber\\
&& \langle f'|\frac{1}{i\eta-H_{0}}H_{A}\frac{1}{i\eta-H_{0}}H_{B}\frac{1}{i\eta-H_{0}}H_{C}|i \rangle\nonumber\\
&& = \sum_{p''\sigma}A_{S1\sigma}\times{A_{S2\sigma}} \\
\end{eqnarray}
Here $A_{S1\sigma}$ and $A_{S2\sigma}$ are given below,
\begin{eqnarray}
A_{S1\sigma}&&=\langle f|H_{DL}\sum_{n=0}^\infty\Big(\frac{1}{i\eta-H_{0}}H_{DL}\Big)^{2n}|f'\rangle \\
A_{S2\sigma}&&=\langle f'|\frac{1}{i\eta-H_{0}}H_{A}\frac{1}{i\eta-H_{0}}H_{B}\frac{1}{i\eta-H_{0}}H_{C}|i\rangle~~~~~~~
\end{eqnarray}

where $|f\rangle = (1/\sqrt{2})(a^{\dagger}_{p\uparrow}a^{\dagger}_{p'\downarrow}- a^{\dagger}_{p\downarrow}a^{\dagger}_{p'\uparrow})|i\rangle$ is the singlet final state and $|f'\rangle = |Dp''\sigma\rangle = d^{\dagger}_{-\sigma}a^{\dagger}_{p''\sigma}|i\rangle$ is the intermediate state whcih tells that one electron is in the dot while the other has tunneled to lead.  The subscripts A, B, C used above in $H$ depends a combination of different tunneling Hamiltonian. In the first case (case (I)) $A,B,C$  denotes  $SD, DL \& SD$ respectively. For the second case (case(I)) $A, B, C$ denote $DL, SD, \& SD$ respectively.  For case (I), as evident, one electron from superconductor  enters into a quantum dot (by the action of $T_{SD}$), the same electron then moves to lead by $T_{DL}$ and finally the second electron arrives at quantum dot by $T_{SD}$. For case (II), it is evident that both the electrons arrive at the same dot by repeated action of $T_{SD}$ costing an interaction energy U. Utilizing the above facts we can simplify as before to abtain $AD_1$ and $AD_2$ in Eq. \ref{fti} and \ref{pqdd}. A straightforward procedure yields the following the expressions for $A_{S\sigma}$, 
\begin{eqnarray}
A_{S1\sigma} && = r_{\sigma}\frac{T_{DL}}{\sqrt{2}}\frac{\epsilon_l+\epsilon_p''-i\eta}{\epsilon_l+\epsilon_p''-i\gamma_l/2}(\delta_{p''p'}+\delta_{p''p}) \\
A_{S2\sigma} && = \frac{T_{DL}T_{SD}^2}{\epsilon_{l}+\epsilon_{p''}-i\eta}\mathcal{B}_{k\sigma}
\end{eqnarray}

In the above $\sigma$ refers spin indices and could be $\uparrow$ or $\downarrow$ and we have  $r_{\uparrow}=-1,~r_{\downarrow}=1$. The expressions for $\mathcal{B}_{k\uparrow}$ and $\mathcal{B}_{k\downarrow}$ are obtained as follows,


\begin{eqnarray}
\mathcal{B}_{k\uparrow} && = \frac{u^c_{1,k}u^c_{2,k} + u^f_{1,k}u^f_{2,k}}{\mathcal{E}_{kq+}^2} + \frac{u^c_{3,k}u^c_{4,k}+u^f_{3,k}u^f_{4,k} }{\mathcal{E}_{kq-}^2}~~~~   \\
\mathcal{B}_{k\downarrow} && = \frac{u^c_{1,\tilde{k}}u^c_{2,\tilde{k}}+u^f_{1,\tilde{k}}u^f_{2,\tilde{k}} }{\mathcal{E}_{kq+}^2} +\frac{u^f_{3,\tilde{k}}u^f_{4,\tilde{k}}+u^f_{3,\tilde{k}}u^f_{4,\tilde{k}} }{\mathcal{E}_{kq-}^2}~~~~
\end{eqnarray}

In the last equation we have used $\tilde{k}=-k$. After some algebra we arrive at the final current expressions as given in Eq. \ref{fincurrent} and Eq. \ref{isamedot}.
Now let's calculate $I_{1}/I_{2}$ for getting an idea about their relative strength. Firstly  for Breit-Wigner resonance we have $\epsilon_1 = -\epsilon_2$,which further reduces $\mathcal{I}_1= \frac{e\gamma_{s}^2}{\pi^2\nu_{s}^2\gamma}\mathcal{I}_D$.
using this we get,
\begin{eqnarray}
\frac{\mathcal{I}_1}{\mathcal{I}_2} = \frac{\mathcal{I}_D}{2\gamma^2{\mathcal{I}_S}}
\end{eqnarray}
Clearly $1/\gamma^2>\mathcal{I}_{S}/\mathcal{I}_{D}$ is the the desired regime where $\mathcal{I}_1$ dominates $\mathcal{I}_2$. We notice that in $\mathcal{I}_S$ expression there is an additional supression of $\mathcal{E}_{kq\pm}$ comapred to $\mathcal{I}_D$ which suggests that $\mathcal{I}_2$ is supressed compared to $\mathcal{I}_1$.

\section{Appendixes-B}
After diagonalisation and filling the negative energy states we obtain the following ground state energy,
The ground state energy of the superconductor is given below, 
\begin{eqnarray}
E_{min}&&=- \sum_{kk's}(\mathcal{E}_{kq+} + \mathcal{E}_{kq-})  -2\frac{\Delta_c \Delta_f}{V_{SC}}+2\frac{m^2}{V_{SD}}~~~~~
\end{eqnarray}
Where  $\mathcal{E}_{kq\pm}= \frac{1}{\sqrt{2}}\sqrt{(\mathcal{P}_k \pm \sqrt{\mathcal{Q}_k})}$. The function $\mathcal{P}_k= \Delta_{c}^2+\Delta_{f}^2+\epsilon_{c}^2(k)+\epsilon_{f}^2(k) +2 m^2$  and  $\mathcal{Q}_k= (\Delta_{c}^2-\Delta_{f}^2+\epsilon_{c}^2(k)-\epsilon_{f}^2(k))^2 + 4m^2((\Delta_{c}-\Delta_{f})^2+(\epsilon_{c}(k)+\epsilon_{f}(k))^2)$.
The Bogoliubov transformations for the present case can be written as follows.

\begin{eqnarray}
\label{bogo1}
c_{k\sigma}&=&u_{1k}^c\gamma_{kq\sigma}^{c}+u_{1\tilde{k}}^c\gamma_{\tilde{k}\tilde{\sigma}}^{{c}\dagger} +u_{1k'}^f\gamma_{k'\sigma}^{f}+u_{1\tilde{k}'}^f\gamma_{\tilde{k}'\tilde{\sigma}}^{{f}\dagger}~~~~\\
\label{bogo2}
c_{\tilde{k}\tilde{\sigma}}^{\dagger}&=&u_{2k}^c\gamma_{k\sigma}^{c}+u_{2\tilde{k}}^c\gamma_{\tilde{k}\tilde{\sigma}}^{{c}\dagger} +u_{2k'}^f\gamma_{k'\sigma}^{f}+u_{2\tilde{k}'}^f\gamma_{\tilde{k}'\tilde{\sigma}}^{{f}\dagger}~~~~\\
\label{bogo3}
f_{k'\sigma}&=&u_{3k}^c\gamma_{k\sigma}^{c}+u_{3\tilde{k}}^c\gamma_{\tilde{k}\tilde{\sigma}}^{{c}\dagger} +u_{3k'}^f\gamma_{k'\sigma}^{f}+u_{3\tilde{k}'}^f\gamma_{\tilde{k}'\tilde{\sigma}}^{{f}\dagger}~~~~\\
\label{bogo4}
f_{\tilde{k}'\tilde{\sigma}}^{\dagger}&=&u_{4k}^c\gamma_{k\sigma}^{c}+u_{4\tilde{k}}\gamma_{\tilde{k}\tilde{\sigma}}^{{c}\dagger} +u_{4k'}\gamma_{k'\sigma}^{f}+u_{4\tilde{k}'}\gamma_{\tilde{k}'\tilde{\sigma}}^{{f}\dagger}~~~~
 \end{eqnarray}

In the above we have used $(\tilde{k}=-k,~\tilde{\sigma}=-\sigma)$.
Also $k'=k+q, \tilde{k}^{\prime}=-k +q$; $q$ is the ordering momentum and  $u$'s are the elements of the eigen vectors of the Hamiltonian given in Eq. 12. Considering '$c$' or '$f$' to be generally denoted by '$p$' and $+k$ or $-k$ denoted by '$z$', the expressions for all the $u$'s can be written as $u_{iz}^p =\frac{a_{iz}^p}{\sqrt{\sum_{zp}(a_{iz}^p)^2}}$. 
Henceforth for simplification we remove the explicit mention of momentum indices in all the variables and  define $\zeta=(\epsilon_f - \epsilon_c)/2, \delta=(\epsilon_f + \epsilon_c)/2$, $\Delta_{\pm}=(\Delta_{c}\pm \Delta_{f})/2$ and $\alpha = \sqrt{(\Delta_{+}\Delta_{-}-\delta\zeta)^2+m^2(\Delta_{-}^2+\delta^2)}$, $ \Omega_{\pm}=\Delta_+\delta+\Delta_-(\zeta \pm E_1)$ and $ \Omega^{\prime}_{\pm}=\Delta_+\delta+\Delta_-(\zeta \pm E_2)$ with $E_{1,2}=\sqrt{A\pm \sqrt{B}}$. Below we provide details expressions of $a_i$'s.
\begin{eqnarray}
 a_{1(2),k}^c &&= \frac{(m^2\delta+(\delta+\zeta+(-)E_1)(\lambda_1-\alpha))}{m\Omega_{+(-)}}, \\
 &&\nonumber \\
 a_{3(4),k}^c &&= \frac{(m^2\delta+(\delta+\zeta+(-)E_2)(\lambda_1+\alpha))}{m\Omega'_{+(-)}}, \\
 &&\nonumber \\
 a_{1(2),{-k}}^c&&=\frac{(\Delta_+(\Delta_-^2+\delta\zeta-\alpha)-\Delta_-(\lambda_2-\alpha))}{m\Omega_{+(-)}}\\
 &&\nonumber \\
 a_{3(4),{-k}}^c&&= \frac{(\Delta_+(\Delta_-^2+\delta\zeta+\alpha)-\Delta_-(\lambda_2+\alpha))}{m\Omega'_{+(-)}}
\end{eqnarray}
In the above $\lambda_1=-\Delta_+\Delta_-+\delta\zeta, \lambda_2=\Delta_+^2+\delta\zeta + m^2$.
Expressions for $a^f_{i,k}$ are comparatively simpler and given by,
\begin{eqnarray}
  a_{1(2),{k}}^f&&=\frac{(\Delta_-^2+\delta^2-\alpha+(-)\delta{E_1})}{\Omega_{+(-)}}\\
 &&\nonumber \\
  a_{3(4),{k}}^f&&=\frac{(\Delta_-^2+\delta^2+\alpha+(-)\delta{E_2})}{\Omega'_{+(-)}}\\
 a_{i,{-k}}^f&&=1,~~i=1,2,3,4
\end{eqnarray}
One can check that the above expressions get simplified for $m=0$ case and we only mention the non-zero elements
below.
\begin{eqnarray}
 a_{3(4),k}^c &&=\frac{-\delta-\zeta-(+)\sqrt{(\Delta_+-\Delta_-)^2+(\delta+\zeta)^2}}{\Delta_--\Delta_+}~~~~\\
a_{1(2),{k}}^f&&=\frac{\delta-\zeta+(-)\sqrt{(\Delta_++\Delta_-)^2+(\delta-\zeta)^2}}{\Delta_-+\Delta_+}~~~~\\
 &&a_{3,{-k}}^c=a_{4,{-k}}^c=a_{1,{-k}}^f=a_{2,{-k}}^f=1
\end{eqnarray}

\end{document}